\documentclass[
aps,pre,onecolumn,superscriptaddress,floatfix,longbibliography]{revtex4-2}

\usepackage{graphicx,graphics,color}
\usepackage[margin=1in]{geometry}
\usepackage{amsmath,amssymb,enumitem,mathtools,ulem}
\usepackage{subcaption}
\usepackage[justification=raggedright,singlelinecheck=false]{caption}
\usepackage[dvipsnames]{xcolor}
\usepackage{float}
\usepackage[hidelinks]{hyperref}

\begin{document}


\title[Preprint]{Effects of Model Reduction on Coherence and Information Transfer in Stochastic Biochemical Systems}

\author{Juan David Marmolejo Lozano}
\affiliation{ 
School of Biological Sciences, University of Edinburgh, Edinburgh, U. K. 
}
\author{Nikola Popovi\'c}
\affiliation{ 
School of Mathematics, University of Edinburgh, Edinburgh, U. K.
}
\affiliation{Maxwell Institute for Mathematical Sciences, University of Edinburgh, Edinburgh, U. K.}
\author{Ramon Grima}
\email{ramon.grima@ed.ac.uk}
\affiliation{ 
School of Biological Sciences, University of Edinburgh, Edinburgh, U. K.
}

\date{\today}

\begin{abstract}
Simplified stochastic models are widely used in the study of frequency-resolved noise propagation in biochemical reaction networks, a common measure being the coherence between random fluctuations in molecule number trajectories. Such models have also found widespread application in the quantification of how information is transmitted in reaction networks via the mutual information (MI) rate. A common assumption is that, under timescale separation, estimates for the coherence and MI rate obtained from simplified (reduced) models closely approximate those in the underlying full models. Here, we challenge that assumption by showing that, while reduced models can faithfully reproduce low-order statistics of molecular counts, they frequently incur substantial discrepancies in the coherence spectrum, especially at intermediate and high frequencies. These errors, in turn, lead to significant inaccuracies in the resulting estimates for the MI rates. We show that the observed discrepancies are due to the interplay between the structure of the underlying reaction networks, the specific model reduction method that is applied, and the asymptotic limits relating the full and the reduced models. We illustrate our results in canonical models of enzyme catalysis and gene expression, highlighting practical implications for quantifying information flow in cells.
\end{abstract}

\maketitle

\section{Introduction}

Realistic models of stochastic biochemical systems are typically difficult to study analytically, or even through stochastic simulation, due to the large number of interacting species and the myriad interactions involved. A much more common approach relies on the construction and study of reduced models that capture the essential dynamics of the underlying ``full" systems, but that involve reduced numbers of interacting species or simplified reaction kinetics. Several methodologies have been devised to achieve model reduction, of which three are commonly used: (i) reduction through separation of timescales \cite{thomas2012rigorous,kim2014validity,song2021universally,popovic2016geometric,kang2019quasi,thomas2014phenotypic}; (ii) reduction through abundance-scale separation \cite{jahnke2011reduced,smith2015model,hellander2007hybrid,cao2018linear,jia2024holimap}; and (iii) reduction by means of effective time-delayed reactions \cite{barrio2013reduction,leier2014exact,zhang2019markovian,szavits2024solving}. Some of these methodologies are equivalent to minimizing the Kullback-Leibler divergence between the full and the reduced models, defined on the space of Monte Carlo simulation trajectories \cite{ocal2023model}. Of the three types of model reduction, the most widely used are methods based on timescale separation; henceforth, we thus exclusively focus on that type.

Reduced models of timescale separation type can be formally proven to estimate accurately molecule number statistics, such as moments and distributions, in subregions of parameter space. A prominent example is the slow-scale Linear-Noise Approximation (ssLNA), the unique reduced form of the standard Linear-Noise Approximation (LNA) of a complex biochemical system under timescale separation conditions, which can be rigorously derived using projection operator techniques \cite{thomas2012rigorous}. Specifically, the ssLNA leads to an accurate reduced model when perturbations about the steady-state concentrations of a subset of the species --- the slow species --- decay on a much slower timescale than those about the steady-state concentrations of the remaining (fast) species; thus, the ssLNA is the stochastic equivalent of the deterministic quasi-steady-state approximation (QSSA). The large-molecule-number limits of several common stochastic reduction methods which assume that the set of governing reactions can be grouped into subsets of slow and fast reactions \cite{cao2005slow,goutsias2005quasiequilibrium,rao2003stochastic} can be shown to be special cases of the ssLNA \cite{thomas2012rigorous}. The accuracy of the ssLNA in calculating the first and second moments of molecular counts, and their power spectra, has been verified by comparison with stochastic simulation via Gillespie's exact algorithm \cite{gillespie1977exact} for a large variety of reaction networks; examples include reactions catalyzed by enzymes, genetic feedback loops, and biologically detailed models of stochastic gene expression \cite{thomas2012slow,ma2024analysis}. An alternative heuristic type of model reduction which is also commonly found in the literature \cite{elf2003fast,ziv2007optimal,rodriguez2010messenger} assumes the form of the probabilities for the effective reactions by comparison with the well-known effective macroscopic rate laws, such as in the assumption of Hill-type propensities in the Chemical Master Equation (CME). The heuristic CME and the associated heuristic LNA, derived from such heuristic propensities, are not generally accurate when timescale separation holds \cite{thomas2012slow,holehouse2019revisiting}.    

Reduced models have also been used to calculate the mutual information (MI) rate between molecule number trajectories of a pair of species in a biochemical reaction network \cite{tostevin2009mutual}. Note that the MI rate is the time-averaged mutual information between a pair of stationary stochastic processes, which is different from instantaneous mutual information; for biochemical networks, the former is more relevant, as the latter cannot correctly quantify the amount of information transmitted per unit time \cite{tostevin2010mutual,meijers2021behavior,fan2024characterizing}. A related concept is the coherence between molecule number trajectories of a pair of species, which measures how strongly the two trajectories are related at each frequency --- within the linear-noise or Gaussian approximation, the coherence completely determines the MI rate via a simple integral formula \cite{munakata2006stochastic,tostevin2009mutual,tostevin2010mutual,komaee2020mutual}. For non-Gaussian fluctuations, as is typically the case, that formula, while not exact, is still useful as a lower bound for the true MI rate \cite{mitra2001nonlinear}. However, the relationship between the coherence and the MI rate predicted by full and reduced models has not been significantly explored, in contrast to the relationship between the moments of the molecular counts, as previously discussed. Since expressions for these quantities have been calculated exclusively for very simple (reduced) models of biological systems, it has in effect been implicitly assumed that the coherence and information transfer measures associated to the reduced models converge to those in the underlying full models in some kinetic parameter limit. For instance, the study of the MI rate between species $X$ and $Y$ interacting via the reaction scheme $\emptyset \rightarrow X \rightarrow \emptyset,~ X \rightarrow X + Y, ~ Y \rightarrow \emptyset$ in \cite{tostevin2009mutual} can be interpreted as a simple model for the transcription of mRNA $X$ and its subsequent translation into protein $Y$; however, clearly that is a reduced model, as many of the biologically important intermediate steps --- such as gene state switching, nuclear export, and binding of mRNA to ribosomes --- are omitted with the implicit understanding that these details are not important to the calculation of the MI rate, provided the intermediate steps are very fast.  

The ssLNA lends itself perfectly to the task of understanding the relationship between the coherence and the MI rate in complex biochemical systems and their corresponding reduced models, as it allows for the straightforward calculation of these measures. Furthermore, the parameter limits relating the full and reduced models can be derived precisely, and the convergence of molecular moments is guaranteed in these limits. In this paper, we demonstrate that, for many reaction networks, model reduction can produce markedly different coherence and MI rate estimates compared to the underlying full models, even when kinetic parameters are chosen so that the full and reduced models yield almost indistinguishable molecular moments at all times.

The paper is organized as follows. In Section II, we provide a concise summary of the general matrix formulation of the LNA, and we define the coherence and the MI rate in that context. In Section III, we summarize the two main types of timescale-separation-based model reduction: a popular heuristic LNA approach and the rigorous ssLNA. In Section IV, we prove that the coherence at zero frequency is exactly predicted by the ssLNA, but not by the heuristic LNA; we also show that the stoichiometry of the reaction networks described by the full and reduced models determines whether the coherence spectra of the two models obey similar or different scaling laws at large frequencies. An important implication is that certain reduced models exhibit infinite MI rates, while those in the underlying full models are finite. In Section V, we illustrate our results in three reaction networks, including in models of enzyme catalysis and gene
expression; furthermore, we identify special systems where model reduction perfectly preserves the coherence and information transfer measures. We conclude with a discussion of our results in Section VI. 


\section{Calculation of coherence and MI rate from the LNA}\label{LNA_MI}

Consider a general chemical reaction network involving $N$ species that are reacting via $R$ reactions:
\begin{align}
    \sum_{i=1}^{N} s_{ij} X_i \xrightarrow{k_j}{}  \sum_{i=1}^{N} r_{ij} X_i, \quad j = 1,\dots, R.
    \label{r_scheme}
\end{align}
Here, $j$ is an index identifying the $j$-th reaction; $X_i$ denotes species
$i$; $s_{ij}$ and $r_{ij}$ are the integer stoichiometric coefficients; and $k_j$ is the macroscopic rate constant of reaction $j$. By the law of mass action, the rate at which reaction $j$ occurs is given by $f_j = k_j \prod_{i=1}^N \phi_i^{s_{ij}}$, where $\phi_i$ is the concentration of species $X_i$. The net change in the number of molecules of $X_i$ when reaction $j$ occurs is $S_{ij} = r_{ij} - s_{ij}$.  

It is well known that in the limit of large volumes, the dynamics of the mean molecule numbers are given by the deterministic rate equations
\begin{align}
\frac{d \vec{\phi}}{d t} = \underline{S} \vec{f}(\vec{\phi},t),
\label{REq}
\end{align}
where $\underline{S}$ is the stoichiometric matrix with elements $S_{ij}$, with $i=1,\dots,N$ and $j=1,\dots,R$; $\vec{f}$ is the rate function vector with elements $f_i$, where $i = 1,\dots,N$; and $\vec{\phi}$ is the mean concentration vector with elements $\phi_i$, with $i=1,\dots,N$.

We define the Jacobian matrix $\underline{J}$ of the rate equations with elements 
\begin{align}
  J_{ij} = \frac{\partial}{\partial \phi_j} \sum_{r = 1}^R S_{ir} f_r  
  \label{Jac}
\end{align}
and the diffusion matrix $\underline{D}$ with elements 
\begin{align}
    D_{ij} = \sum_{r=1}^R S_{ir} S_{jr} f_r,
    \label{dmatrix}
\end{align}
which can be compactly written as $\underline{D}=\underline{S}\, \underline{F}\, \underline{S}^T$, where $\underline{F}$ is a diagonal matrix whose elements are $f_i$ ($i=1,\dots,N$).

Note that $D_{ii} > 0$, but that $D_{ij}\ (i \ne j)$ is not sign-definite and that it can also equal zero. Note also that the diffusion matrix $D$ is positive semi-definite, which means that it is symmetric $(D_{ij} = D_{ji})$, and that it obeys the inequality $D_{ij}^2 \le D_{ii}D_{jj}$ \cite{van1992stochastic,horn2012matrix}.
In what follows, we will assume steady-state conditions: the time derivative in Eq.~\eqref{REq} is set to zero, which implies that the concentrations and the Jacobian and diffusion matrices are time-independent functions of the rate constants. 

In the limit of large volume $\Omega$ and under the assumption that the system of rate equations in Eq.~\eqref{REq} has one unique steady state, fluctuations of molecule numbers are described by an Ornstein-Uhlenbeck process that is commonly called the Linear-Noise Approximation (LNA) \cite{van1992stochastic,schnoerr2017approximation}. Under that approximation, the variance-covariance matrix $\underline{C}$ of the concentration fluctuations is given by the solution of the Lyapunov equation
\begin{align}
    \frac{d \underline{C}}{d t} = \underline{J}\,\underline{C} + \underline{C}\,\underline{J}^T + \frac{\underline{D}}{\Omega};
    \label{Leq}
\end{align}
this matrix formulation of the LNA is originally due to Elf and Ehrenberg \cite{elf2003fast}. The power spectrum matrix for the fluctuations in concentration is given by
\begin{align}
    \underline{P} (\omega) = \frac{1}{\Omega} (\underline{J} + i\underline{I}_N\omega)^{-1} \underline{D}(\underline{J}^T - i\underline{I}_N\omega)^{-1}, 
    \label{ps}
\end{align}
where $\underline{I}_N$ is the identity matrix of rank $N$ and $\omega$ is frequency \cite{gardiner1985handbook,thomas2012slow}. The (magnitude-squared) coherence $\xi_{ij}$ between the timeseries of the fluctuations of the numbers of molecules of species $X_i$ and $X_j$ is given by
\begin{align}
    \xi_{ij}(\omega) = \frac{|P_{ij}(\omega)|^2}{P_{ii}(\omega) P_{jj}(\omega)}.
    \label{Coh}
\end{align}
Since both factors in the denominator of Eq.~\eqref{Coh} are positive and the numerator is non-negative for any $\omega$, it follows that the coherence is non-negative.  
By the Cauchy-Schwarz inequality, one can also show that $\xi_{ij}(\omega) \le 1$. 

Note that the coherence can be interpreted as a correlation analysis in the frequency domain, since $\xi_{ij}(\omega)$ equals the square of the  correlation coefficient of the frequency component $\omega$ of the timeseries of $X_i$ and $X_j$. Hence, $\xi_{ij}(\omega) = 0$ when the fluctuations in $X_i$ and $X_j$ at frequency $\omega$ are uncorrelated, whereas $\xi_{ij}(\omega) = 1$ when they are perfectly correlated. That interpretation is due to the Wiener-Khinchin Theorem: the variance is the integral of the power spectrum over $\omega$, while the covariance is the integral of the cross-power spectrum over $\omega$,  
\begin{align}
    \xi_{ij}(\omega) = \frac{|P_{ij}(\omega)\Delta \omega|^2}{(P_{ii}(\omega)\Delta \omega)(P_{jj}(\omega)\Delta \omega)} = \frac{C_{ij}(\omega)^2}{C_{ii}(\omega)C_{jj}(\omega)},
    \label{Corr}
\end{align}
where $C_{ij}(\omega)$ is the covariance of the two signals after they are filtered by passing through a bandpass filter centered at frequency $\omega$ and infinitesimally small bandwidth $\Delta \omega$; similarly, $C_{ii}(\omega)$ and $ C_{jj}(\omega)$ are the variances of the signals $X_i$ and $X_j$ after filtering. In what follows, we shall refer to a plot of the coherence as a function of frequency as the coherence spectrum. 

    \label{avcoh}
   \label{weightedcoh}

Since the fluctuations in the LNA are Gaussian distributed, it also follows that the MI rate between the two timeseries is given by 
\begin{align}
    M_{ij} = -\frac{1}{4 \pi \ln 2} \int_{-\infty}^{\infty} \text{ln}(1 - \xi_{ij}(\omega)) d\omega.  
    \label{MIrate}
\end{align}
Because of the factor of $\ln 2$, the MI rate is expressed in units of bits per unit time. Note that the formula for the MI rate in Eq.~\eqref{MIrate} stems from its formal definition as the limit of the time-averaged mutual information between corresponding segments of two stationary stochastic processes, as the segment length approaches infinity \cite{tostevin2009mutual,komaee2020mutual}. It can also be proven analytically that Eq.~\eqref{MIrate} provides a lower bound on the MI rate when the fluctuations are non-Gaussian \cite{mitra2001nonlinear} which might not be tight \cite{reinhardt2023path,moor2023dynamic}. Since, in reality, molecular fluctuations in biochemical systems are discrete, they are never truly Gaussian as assumed by the LNA; hence, we shall purely interpret Eq.~\eqref{MIrate} as an analytically tractable lower bound for the actual MI rate, which can be estimated, in some cases, by Monte Carlo approaches based on path weight sampling \cite{reinhardt2023path}.     

\section{Reduced LNA under timescale separation conditions: heuristic and rigorous approaches}

Consider the case where a subset of the $N$ species interacting via the reactions in Eq.~\eqref{r_scheme} are slow, while the remainder are fast. Let $N_s$ and $N_f$ be nonnegative integers with $N_s+N_f=N$, where $N_s$ and $N_f$ are the numbers of slow and fast species, respectively. Specifically, fast species exhibit rapid relaxation of small deviations from their steady-state concentrations, while slow species relax on much longer timescales.

\subsection{Heuristic model reduction} \label{heurmodelred}

A heuristic reduction method is based on the deterministic QSSA: in the rate equations, the time derivatives of the concentrations of the fast species are set to zero, and expressions are derived for the concentrations of the fast species in terms of the slow species. These expressions are then substituted into the rate equations for the slow species to obtain a set of new rate equations that are specifically functions of the slow species only. 

A common example of this type of reduction occurs in the modeling of the enzyme-catalyzed reaction network, $S + E\xrightleftharpoons[]{}C\xrightarrow{}E+P$, where $S$ is substrate, $E$ is enzyme, $C$ is complex, and $P$ is product. Under the assumption that $E$ and $C$ are the fast species and $S$ and $P$ are the slow species, the QSSA allows for the rate equations to be reduced to those describing the simpler system $S \xrightarrow{} P$ with an effective non-linear rate proportional to $\phi_S/(K_M + \phi_S)$, where $\phi_S$ is the substrate concentration and $K_M$ denotes the Michaelis-Menten constant.

For a general system that is reduced via the QSSA, to the reduced rate equations there corresponds a coarse-grained reaction network composed of $R^\ast < R$ effective reactions between the slow species: 
\begin{align}
    \sum_{i=1}^{N_s} s_{ij}^{*} X_i^s \rightarrow  \sum_{i=1}^{N_s} r_{ij}^{*} X_i^s, \quad j = 1,\dots, R^\ast.
    \label{r_scheme_red}
\end{align}
Here, $X_i^s$ denotes the $i$-th slow species, and $s_{ij}^*$ and $r_{ij}^*$ are the integer stoichiometric coefficients. The rate at which reaction $j$ occurs is some function $f_j^* (\phi_1^{s},\dots,\phi_{N_s}^{s})$, where $\phi_i^s$ is the concentration of the slow species $X_i^s$. Note that the function $f_j^\ast$ is not generally consistent with the law of mass action, an example being the Michaelis-Menten rate law discussed above, as it describes the rate of an effective reaction, not an elementary one. The net change in the number of molecules of $X_i^s$ when reaction $j$ occurs is $S_{ij}^\ast = r_{ij}^\ast - s_{ij}^\ast$. 

The LNA of this effective reduced system of reactions is given by Eqs.~\eqref{Jac} through \eqref{ps}, with $S_{ij}$ and $f_j$ replaced by $S_{ij}^\ast$ and $f_j^\ast$, respectively. The coherence then follows directly from Eq.~\eqref{Coh}. The resulting method for constructing a reduced LNA is heuristic, because (i) it is not rigorously derived from the LNA of the full system under timescale separation conditions; and (ii) it assumes that the reduced microscopic kinetics is similar to the reduced deterministic and macroscopic kinetics. The heuristic reduction is one the methods we consider in this paper, as it has been widely applied, including in recent studies of the MI rate \cite{moor2023dynamic,reinhardt2025mutual}.  

\subsection{Rigorous model reduction}\label{rigorousred}

In \cite{thomas2012rigorous,thomas2012slow}, a rigorous adiabatic elimination method was developed for the LNA. Starting from the Fokker-Planck equation describing the LNA of the full system, the method, which is called the ssLNA, allows for the derivation of a simplified Fokker-Planck equation that describes the time evolution of the joint probability density function of the slow variables only. In what follows, we briefly summarize the method -- we shall refer to the system of fast and slow interacting species as the full model, while we denote the reduced system of slow species interacting via effective reactions as the reduced model.

First, we relabel the species in Eq.~\eqref{r_scheme} so that $X_1,\dots,X_{N_s}$ are the slow ones and $X_{N_s+1},\dots,X_{N_f}$ are the fast ones. We partition the $N\times N$ square matrix $\underline{J}$ whose elements are given by Eq.~\eqref{Jac} and the $N\times R$ matrix $\underline{S}$ as
\[
\underline{J}=\begin{pmatrix}\underline{J}_{ss} & \underline{J}_{sf}\\[4pt] \underline{J}_{fs} & \underline{J}_{ff} \end{pmatrix}\quad\text{and}\quad
\underline{S}=\begin{pmatrix} \underline{S}_s\\[4pt] \underline{S}_f\end{pmatrix},
\]
respectively. Here, the dimensions of the rectangular (or square) sub-matrices $\underline{J}_{ss}$, $\underline{J}_{sf}$, $\underline{J}_{fs}$, and $\underline{J}_{ff}$ are $N_s\times N_s$, $N_s\times N_f$, $N_f\times N_s$, and $N_f\times N_f$, respectively, while $\underline{S}_{s}$ and $\underline{S}_{f}$ have dimensions $N_s\times R$ and $N_f\times R$, respectively. 

The effective Jacobian and diffusion matrices, valid under timescale separation conditions, are given by
\begin{align}
    \underline{J}_S&= \underline{J}_{ss} - \underline{J}_{sf}\underline{J}_{ff}^{-1}\underline{J}_{fs}, \label{JSdef} \\
    \underline{D}_S &= (\underline{A}-\underline{B})(\underline{A}-\underline{B})^T,
    \label{effD}
\end{align}
where
\[
\underline{A} = \underline{S}_s\sqrt{\underline{F}}\quad\text{and}\quad \underline{B} = \underline{J}_{sf}\underline{J}_{ff}^{-1} \underline{S}_f \sqrt{\underline{F}}.
\]
Note that $\underline{J}_S$ is the Schur complement of the sub-matrix $\underline{J}_{ff}$ of the matrix $\underline{J}$, which is defined provided $\underline{J}_{ff}$ is invertible. Note also that $\underline{F}$ is the $R \times R$ diagonal matrix with elements of $\vec{f}$, i.e. of the rate function vector in the full model. 

Next, Eq.~\eqref{effD} can be written in the form $\underline{D}=\underline{S}' \underline{F}\, \underline{S}'^T$, where $\underline{S}'$ is an effective stoichiometric matrix of the reduced system given by
\begin{align}
\underline{S}' = \underline{S}_s - \underline{J}_{sf} \underline{J}_{ff}^{-1} \underline{S}_f.    
\label{Sprime}
\end{align}
Notably, the entries of $\underline{S}'$ are generally not integers, which reflects the fact that reduction of a full model described by the CME does not typically result in a model that can be described by a different CME, as the reduced dynamics may be non-Markovian \cite{janssen1989elimination} --- a Langevin formulation of the reduced ssLNA dynamics always exists, however.

In analogy to Eqs.~\eqref{Leq} and \eqref{ps}, the equations for the variance-covariance matrix $\underline{C}_S$ of concentration fluctuations of the slow species obtained from the ssLNA read
\begin{align}
    \frac{d \underline{C}_S}{d t} = \underline{J}_S \underline{C}_S + \underline{C}_S \underline{J}_S^T + \frac{\underline{D}_S}{\Omega},
    \label{Leq_slow}
\end{align}
while the power spectrum matrix $\underline{P}_S$ for the concentration fluctuations is given by
\begin{align}
    \underline{P}_S (\omega) = \frac{1}{\Omega} (\underline{J}_S + i\underline{I}_{N_s}\omega)^{-1} \underline{D}_S(\underline{J}_S^T - i\underline{I}_{N_s}\omega)^{-1}, 
    \label{ps_slow}
\end{align}
where $\underline{I}_{N_s}$ is the identity matrix of rank $N_s$ and $\omega$ is frequency. The coherence between any two slow species can then be calculated from Eq.~\eqref{Coh}, with $P_{ij}(\omega)$ replaced by $\big[\underline{P}_S\big]_{ij}(\omega)$.

We note that, generally, the ssLNA does not correspond to the heuristic LNA discussed earlier. The heuristic approach has been shown to be equivalent to the implicit assumption that under timescale separation conditions, reversible elementary reactions involving fast species do not contribute to the intrinsic noise in the concentrations of slow species -- an assumption that is generally incorrect \cite{thomas2012slow}. It can be shown that heuristic model reduction and the ssLNA always result in the same set of reduced deterministic rate equations and, hence, to the same Jacobian $\underline{J}_S$; however, they differ in their diffusion matrices.

\section{General relationships between coherence spectra and MI rates of the full and reduced models}

\subsection{Coherence in the limit of zero frequency} \label{IVA}

In this section, we first establish a relationship between the predictions of the full model corresponding to the full LNA and the reduced models obtained from the ssLNA and the heuristic LNA for the power spectra and the cross-power spectra of the slow species at $\omega = 0$. Subsequently, we apply these results to investigate the relationship between the coherences of any pair of slow species at $\omega = 0$, as predicted by the full and reduced models. 

From Eqs.~\eqref{ps} and \eqref{ps_slow}, it follows that the spectra at $\omega = 0$ of the full LNA and the ssLNA are given by
\[
P_{ij}(0) = \big[\,\underline{J}^{-1} \underline{D}\, \underline{J}^{-T}\,\big]_{ij}\quad\text{and}\quad
\big[\underline{P}_S\big]_{ij}(0) = \big[\,\underline{J}_S^{-1}\,\underline{D}_S\,\underline{J}_S^{-T}\,\big]_{ij},
\]
respectively, where $i,j\in\{1,\dots,N_s\}$. Note that for the full model, we merely consider the upper left $N_s\times N_s$ block of $\underline{J}^{-1} \underline{D} \underline{J}^{-T}$, as we are only concerned with the spectra of the slow species.

We proceed in two steps. First, we derive the block-inverse formula for $\underline{J}^{-1}$. Second, we evaluate the upper left block of $\underline{J}^{-1}\underline{D}\,\underline{J}^{-T}$ and simplify the resulting expression to show that it coincides exactly with $\underline{J}_S^{-1}\,\underline{D}_S\,\underline{J}_S^{-T}$.

The calculations for Step 1 can be found in Appendix~\ref{Appinv}. The result is a simple formula for the block-inverse of the Jacobian of the full model: 
\[
\underline{J}^{-1} =
\begin{pmatrix}
\underline{J}_S^{-1} & -\,\underline{J}_S^{-1} \underline{J}_{sf}\underline{J}_{ff}^{-1} \\[6pt]
-\,\underline{J}_{ff}^{-1}\underline{J}_{fs} \underline{J}_S^{-1} & \; \underline{J}_{ff}^{-1} + \underline{J}_{ff}^{-1} \underline{J}_{fs} \underline{J}_S^{-1} \underline{J}_{sf} \underline{J}_{ff}^{-1}
\end{pmatrix}.
\;
\]

Next, we proceed to Step 2. Let the upper left block of $\underline{J}^{-1}\underline{D}\,\underline{J}^{-T}$ be denoted by $\underline{P}_{ss}$, and write the matrices in the triple product as
\[
\underline{J}^{-1}=
\begin{pmatrix}
\underline{X} & \underline{Y}\\[4pt]
\underline{Z} & \underline{W}
\end{pmatrix},\quad
\underline{D}=
\begin{pmatrix}
\underline{D}_{ss} & \underline{D}_{sf}\\[4pt]
\underline{D}_{fs} & \underline{D}_{ff}
\end{pmatrix},\quad\text{and}\quad
\underline{J}^{-T}=
\begin{pmatrix}
\underline{X}^T & \underline{Z}^T\\[4pt]
\underline{Y}^T & \underline{W}^T
\end{pmatrix}.
\]
It then follows that only the following terms contribute to the upper left block of the product of these three matrices:
\begin{align}
\underline{P}_{ss}=\underline{X}\,\underline{D}_{ss} \underline{X}^T + \underline{X}\,\underline{D}_{sf} \underline{Y}^T + \underline{Y}\, \underline{D}_{fs} \underline{X}^T + \underline{Y}\,\underline{D}_{ff} \underline{Y}^T.
\label{Pss}
\end{align}

Hence, to evaluate $\underline{P}_{ss}$, we only need the upper left and upper right blocks from the block-inverse formula for $\underline{J}^{-1}$, i.e. $\underline{X} = \underline{J}_S^{-1}$ and $\underline{Y}=-\underline{J}_S^{-1}\underline{J}_{sf}\underline{J}_{ff}^{-1}$. We also require the block elements of $\underline{D}$ --- since $\underline{D} = \underline{S}\, \underline{F}\, \underline{S}^T=\big(\underline{S}\sqrt{\underline{F}}\big)\big(\underline{S}\sqrt{\underline{F}}\big)^T$, its block form is
\[
\underline{D} =
\begin{pmatrix}
\underline{D}_{ss} & \underline{D}_{sf} \\[4pt]
\underline{D}_{fs} & \underline{D}_{ff}
\end{pmatrix}
=
\begin{pmatrix}
\underline{A}\, \underline{A}^{T} & \underline{A}\, \underline{E}_f^{T} \\[4pt]
\underline{E}_f \underline{A}^{T} & \underline{E}_f \underline{E}_f^{T}
\end{pmatrix},
\]
where $\underline{A} = \underline{S}_s\sqrt{\underline{F}}$ and $\underline{E}_f = \underline{S}_f\sqrt{\underline{F}}$.
Substituting these expressions and the block formulas for $\underline{D}$ into Eq.~\eqref{Pss}, we obtain
\[
\begin{aligned}
\underline{P}_{ss}
&= \underline{J}_S^{-1}\,\big(\underline{A}\,\underline{A}^T\big)\,\underline{J}_S^{-T} + \underline{J}_S^{-1}\,\big(\underline{A}\, \underline{E}_f^T\big)\,\big(-\underline{M}^T \underline{J}_S^{-T}\big)\\[4pt]
&\quad + \big(-\underline{J}_S^{-1}\underline{M}\big)\,\big(\underline{E}_f \underline{A}^T\big)\,\underline{J}_S^{-T}
  + \big(-\underline{J}_S^{-1}\underline{M}\big)\,\big(\underline{E}_f \underline{E}_f^T\big)\,\big(-\underline{M}^T \underline{J}_S^{-T}\big),
\end{aligned}
\]
where $\underline{M} = \underline{J}_{sf}\underline{J}_{ff}^{-1}$. Factoring \(\underline{J}_S^{-1}\) on the left and \(\underline{J}_S^{-T}\) on the right, we find
\[
\begin{aligned}
\underline{P}_{ss}
&= \underline{J}_S^{-1}\big[\, \underline{A}\,\underline{A}^T - \underline{A}\, \underline{E}_f^T \underline{M}^T - \underline{M}\, \underline{E}_f \underline{A}^T + \underline{M}\, \underline{E}_f \underline{E}_f^T \underline{M}^T \,\big] \underline{J}_S^{-T}.
\end{aligned}
\]
Now, recall that $\underline{B} = \underline{M}\, \underline{E}_f = \underline{J}_{sf}\underline{J}_{ff}^{-1}\underline{S}_f\sqrt{ \underline{F}}$. It can also be shown using Eq.~\eqref{effD} that the diffusion matrix in the ssLNA is given by
\begin{align*}
\underline{D}_S &= (\underline{A} - \underline{B})(\underline{A} - \underline{B})^T = \underline{A}\,\underline{A}^T - \underline{A}\, \underline{B}^T - \underline{B}\, \underline{A}^T + \underline{B}\, \underline{B}^T \\
&= \underline{A}\,\underline{A}^T - \underline{A}\, \underline{E}_f^T \underline{M}^T - \underline{M}\, \underline{E}_f \underline{A}^T + \underline{M}\, \underline{E}_f \underline{E}_f^T \underline{M}^T.
\end{align*}
Therefore, we have
\[
\underline{P}_{ss} = \underline{J}_S^{-1}\,\underline{D}_S\,\underline{J}_S^{-T}.
\]
Hence, by Eq.~\eqref{ps_slow}, it follows that the power spectra and the cross-spectra of the slow species, as predicted by the full LNA and the ssLNA, coincide at zero frequency, 
\begin{align}
P_{ij}(0) = \big[\underline{P}_S\big]_{ij}(0),
\end{align}
where $i,j\in\{1,\dots,N_s\}$. 

\vspace{0.5cm}
\textit{Conclusion.} By Eq.~\eqref{Coh}, it follows that \textit{the coherences in the full model and the reduced model obtained from the ssLNA are precisely equal at $\omega = 0$}. Since the heuristic model reduction leads to an LNA with the same Jacobian, but a different diffusion matrix, than the ssLNA, it follows that the coherence of the heuristic reduced model cannot generally be guaranteed to be correct at $\omega = 0$. 

\subsection{Coherence spectrum in the asymptotic limit of large frequency} \label{IVB}

We start by considering the leading-order terms for the cross-spectra $P_{ii}$ and spectra $P_{ij}$ of the full model in the limit of large $\omega$. 

Let $\underline{\Lambda} = \underline{J} + i \omega \underline{I}_N$ and $\underline{\Delta} = \underline{J}^T - i \omega \underline{I}_N$. Note that these matrices are key components in the equation for the power spectrum given by Eq.~\eqref{ps},
\begin{equation}
  \underline{\Lambda}^{-1} = (i\omega)^{-1} \Big( \underline{I}_N - i \frac{\underline{J}}{\omega} \Big)^{-1} = (i\omega)^{-1} \sum_{k=0}^\infty i^k \frac{\underline{J}^k}{\omega^k} = \sum_{k=0}^\infty i^{k-1} \frac{\underline{J}^k}{\omega^{k+1}}, 
\end{equation}
where we used the Neumann series representation, which is valid in the limit of large $\omega$, in the last two steps. Similarly, we can write
\begin{equation}
  \underline{\Delta}^{-1} = (-i\omega)^{-1} \bigg( \underline{I}_N + i \frac{\underline{J}^T}{\omega} \bigg)^{-1} = (-i\omega)^{-1} \sum_{k=0}^\infty (-i)^k \frac{\big(\underline{J}^T\big)^k}{\omega^k} = \sum_{k=0}^\infty (-i)^{k-1} \frac{\big(\underline{J}^T\big)^k}{\omega^{k+1}}. 
\end{equation}
Now, by Eq.~\eqref{ps}, we have 
\begin{align}
    \underline{P} &= \Omega^{-1} \underline{\Lambda}^{-1}\underline{D}\,\underline{\Delta}^{-1}, \\
    &= \Omega^{-1} \sum_{k=0}^\infty  \sum_{m=0}^\infty i^{k-1} (-i)^{m-1} \frac{\underline{J}^k}{\omega^{k+1}} \underline{D} \frac{\big(\underline{J}^T\big)^m}{\omega^{m+1}}, \\
    &= \Omega^{-1} \sum_{k=0}^\infty  \sum_{m=0}^\infty \frac{i^{k+3m-4}}{\omega^{k+m+2}} \underline{J}^k \underline{D} \big(\underline{J}^T\big)^m.
    \label{psseries}
\end{align}
In the limit of large $\omega$, the first-order series expansion (in $\omega^{-1}$) for $\underline{P}$ is hence given by
\begin{equation}
    \underline{P} = \frac{\Omega^{-1}\underline{D}}{\omega^2} - \frac{i \Omega^{-1}}{\omega^3}(\underline{D}\,\underline{J}^T - \underline{J}\,\underline{D}) + O(\omega^{-4}),
    \label{pseriesfinal}
\end{equation}
which is obtained for $(k, m) = (0, 0)$, $(k, m) = (0, 1)$, and $(k, m) = (1, 0)$ in Eq.~\eqref{psseries}. For a general system of interacting chemical species, $\underline{D}\,\underline{J}^T \ne \underline{J}\,\underline{D}$, which implies that the second term in Eq.~\eqref{pseriesfinal} is non-zero. For a system of non-interacting species, it is straightforward to show from the definitions of the Jacobian and diffusion matrices in Eqs.~\eqref{Jac} and \eqref{dmatrix} that both $\underline{D}$ and $\underline{J}$ are diagonal matrices and, hence, that $\underline{D}\,\underline{J}^T = \underline{J}\,\underline{D}$. 


Using Eqs.~\eqref{Coh}, \eqref{MIrate}, and \eqref{pseriesfinal}, we find that if $D_{ij} \ne 0$, then
\begin{align}
    \xi_{ij}(\omega)&\rightarrow \frac{D_{ij}^2}{D_{ii}D_{jj}}\quad\text{as }\omega \rightarrow \infty \label{epsij1} 
\end{align}
and, hence, that $M_{ij}\to\infty$. 
Recall that, because the diffusion matrix $\underline{D}$ is symmetric and positive semi-definite \cite{van1992stochastic}, $D_{ij}^2 \le D_{ii}D_{jj}$ holds \cite{horn2012matrix}, which implies $0 \le \xi_{ij}(\omega)\le 1$. 

By contrast, if $D_{ij} = 0$, then we find
\begin{align}
    \xi_{ij}(\omega) &\sim \frac{([\underline{D}\,\underline{J}^T]_{ij} - [\underline{J}\,\underline{D}]_{ij})^2}{D_{ii}D_{jj}} \frac{1}{\omega^2}\quad\text{as }\omega \rightarrow \infty, \label{scaling}
\end{align}
which implies that
\begin{align}
M_{ij}\geq 0\quad\text{is finite.}
\label{finiteint}
\end{align}
Note that, although $\underline{D}\,\underline{J}^T \ne \underline{J}\,\underline{D}$ in general for a system of interacting species, as previously mentioned, it is possible for specific parameter values to yield $[\underline{D}\,\underline{J}^T]_{ij} = [\underline{J}\,\underline{D}]_{ij}$ --- in that case, $\xi_{ij}(\omega) \propto\omega^{-4}$ as $\omega \rightarrow \infty$; see below for an example. A proof of Eq.~\eqref{finiteint} can be found in Appendix~\ref{AppA}. 

Next, we give a physical interpretation of these results. The case where $D_{ij} = \sum_{p=1}^R S_{ip}S_{jp} f_p(\vec{\phi}) \ne 0$ occurs when \textit{in the set of reactions defining the biochemical system, there exists at least one reaction in which the numbers of molecules of species $X_i$ and $X_j$ change simultaneously when the reaction occurs}. For such reaction networks, the LNA predicts an infinite MI rate between species $X_i$ and $X_j$, which reflects the fact that the change in the numbers of molecules of $X_i$ and $X_j$ is \textit{perfectly coupled} in that case; e.g. when the reaction $X_i \rightarrow X_j$ occurs, the number of molecules of species $X_i$ decreases by $-1$, while the number of molecules of species $X_j$ increases by $+1$. 

By contrast, \textit{when the species $X_i$ and $X_j$ are either not involved in a common reaction, or else if they are involved in one or more common reactions such that the net change in the number of molecules of one of the species is zero}, then $D_{ij} = 0$. The reason is that in these cases, for any reaction $p$, either $S_{ip}$ or $S_{jp}$ will be zero. Then, the LNA predicts a finite MI rate, which reflects the imperfect coupling of the fluctuations of the numbers of molecules of species $X_i$ and $X_j$. We note that these conclusions hold for any pair of species -- slow or fast, or a combination thereof -- described by the LNA of the full model.

Starting from Eq.~\eqref{ps_slow}, similar results can be derived for the power spectra and cross-power spectra of slow species from the ssLNA: Eqs.~\eqref{epsij1} and \eqref{scaling} hold, with $\underline{D}$ and $\underline{J}$ replaced by $\underline{D}_S$ and $\underline{J}_S$, respectively. In that case, $[\underline{D}_S]_{ij}=\sum_{p=1}^R S_{ip}'S_{jp}'f_p(\vec{\phi})$, where $\underline{S}'$ is the effective stoichiometric matrix of the reduced system; see Eq.~\eqref{Sprime}. Under timescale separation, the condition that $[\underline{D}_S]_{ij} \ne 0$ holds when there exists at least one reaction in which the numbers of molecules of the slow species $X_i$ and $X_j$ effectively change simultaneously when the reaction occurs.   
\vspace{0.5cm}

\textit{Conclusion.} The coherence spectra of two slow species $X_i$ and $X_j$ in the full model and the reduced model obtained from the ssLNA obey different scaling laws for large $\omega$ if in only one of these two models, there exists at least one reaction in which the numbers of molecules of species $X_i$ and $X_j$ change simultaneously when the reaction occurs --- one of the models then predicts an infinite MI rate, while the MI rate predicted by the other is finite. If both the full model and the ssLNA obey the aforementioned reaction constraint or if neither obeys it, then the coherence spectra satisfy the same scaling laws for large $\omega$; nevertheless, these might still not agree exactly if the corresponding prefactors differ. In that case, both models will predict either an infinite MI rate or a finite MI rate. In summary, model reduction via the ssLNA generally does not preserve the coherence spectrum, except for very small frequencies. The same general conclusions hold for a comparison of the coherence spectra of the slow species in the full model and the heuristic reduced model, except that there is also lack of agreement at $\omega = 0$ then. 

\section{Applications}

\subsection{Michaelis-Menten model of enzyme catalysis with substrate synthesis}

Consider the set of reactions
\begin{align}
    \varnothing \xrightarrow{k_1} S,\quad S + E\xrightleftharpoons[k_3]{k_2}C,\quad C\xrightarrow{k_4}E+P,\quad P\xrightarrow{k_5}\varnothing
    \label{fullenzyme}
\end{align}
describing a conventional enzyme-catalyzed reaction network in which substrate molecules $S$ are produced, bind enzyme molecules $E$ to form complex $C$, and then disassociate into the original unbound enzyme form $E$ and product molecules $P$. The latter are subsequently degraded. Note that the total concentration $\phi_T$ of enzyme molecules in unbound and bound form is a time-independent constant, since enzyme only promotes the reaction but is itself neither produced nor destroyed. We do not explicitly model substrate decay as the corresponding rate, which for most proteins is typically due to dilution, is much smaller than the effective removal rate of substrate by the enzyme reaction. 

In the large-system-size limit, the mean concentrations of all chemical species in Eq.~\eqref{fullenzyme} are given by the deterministic rate equations in Eq.~\eqref{REq}; solution of these equations at steady state leads to the mean concentrations for substrate, enzyme, and product,
\begin{align}
   \label{solenzy}
   \phi_S = \frac{K_M x}{1 - x},\quad \phi_E = \phi_T - \frac{k_1}{k_4},\quad\text{and}\quad \phi_P = \frac{k_1}{k_5},
\end{align}
where $K_M =(k_3+k_4)/k_2$ is the Michaelis-Menten constant and $x = k_1/k_4 \phi_T$. Note that steady-state conditions only ensue provided that the substrate production rate is lower than the maximum rate at which enzyme can catalyze substrate into product, i.e. for $x < 1$. The free enzyme and complex concentrations are related by a conservation law $\phi_E + \phi_C = \phi_T$, where $\phi_T$ is the total (constant) enzyme concentration. 

From the rate equations, it can be shown (Appendix~\ref{AppB}) that the timescales associated with the various species are given by
\begin{align}
    \tau_S = \frac{1}{k_2 \phi_T} \frac{1}{1 - x}, \quad \tau_P = \frac{1}{k_5},\quad\text{and}\quad\tau_C = \frac{1-x}{k_3+k_4}.
    \label{tscond}
\end{align}
We shall consider the case where substrate and product are the slow species and complex (and free enzyme) are the fast ones. The associated timescale separation condition is $\tau_S, \tau_P \gg \tau_C$. Then, one can reduce the deterministic equations of the full model to those of a simpler system of reactions by means of the QSSA:
\begin{align}
    \varnothing \xrightarrow{k_1} S,\quad S \xrightarrow{k'}P,\quad P\xrightarrow{k_5}\varnothing.
    \label{redenzyme}
\end{align}
Here, the enzyme and complex dynamics are not explicitly modeled; rather, they are implicitly considered through a judicious choice of the effective rate constant $k'$. The rate of the effective reaction $S \rightarrow P$ is of the Michaelis-Menten form $k' \phi_S = k_4 \phi_T \phi_S / (K_M + \phi_S)$. 

We refer to the LNA corresponding to the full circuit in \eqref{fullenzyme} as the full LNA. Similarly, the heuristic LNA is the LNA corresponding to the reduced circuit in \eqref{redenzyme}. A second, more rigorous way to obtain a reduced LNA when complex is fast relies on the ssLNA. The corresponding calculations can be found in Appendix~\ref{AppB}. Therein, we show that, while all three approaches predict the same means for all species, the second moments differ. In particular, the ssLNA agrees with the full LNA whenever the timescale separation condition $\tau_S, \tau_P \gg \tau_C$ is satisfied. However, the heuristic LNA requires an additional constraint, $k_3 \gg k_4$, which we call the ``infrequent catalysis constraint", as it implies that when complex $C$ is formed, it is much more probable that it decays to $E$ and $S$ rather than to $E$ and $P$. Similar results have previously been reported for simpler metabolic systems \cite{thomas2011communication}. 

It can also be shown that the coherences between substrate and product fluctuations obtained from the LNA of the full model, the ssLNA, and the heuristic LNA are given by
\begin{align}
    \xi_{S,P}^{\text{full}}(\omega) &= \frac{c_0 + c_1 \omega^2 + c_2 \omega^4}{c_3 + c_4 \omega^2 + c_5 \omega^4 + c_6 \omega^6},
    \label{cohfullenzyme} \\
    \xi_{S,P}^{\text{ssLNA}}(\omega) &= \frac{d_0 + d_1 \omega^2}{d_2 + d_3 \omega^2},\quad\text{and}
    \label{cohssLNAenzyme} \\
    \xi_{S,P}^{\text{heur}}(\omega) &= \frac{1}{4},
    \label{cohheuristicenzyme}
\end{align}
respectively. Here, $c_i$ ($i=0,1,3,4,5,6$) and $d_i$ ($i=0,\dots,3$) are positive constants, while $c_2$ is a non-negative constant, all of which are functions of the reaction rates $k_i$ and the total enzyme concentration $\phi_T$; for precise formulae, see Appendix~\ref{AppB}. 



Our main question now is: given a choice of parameters such that the ssLNA and the heuristic LNA accurately predict the statistics of substrate and product, does it follow that the corresponding predictions for the coherence spectrum are accurate?  

\begin{figure}
        \centering       
        \includegraphics[scale=0.7]{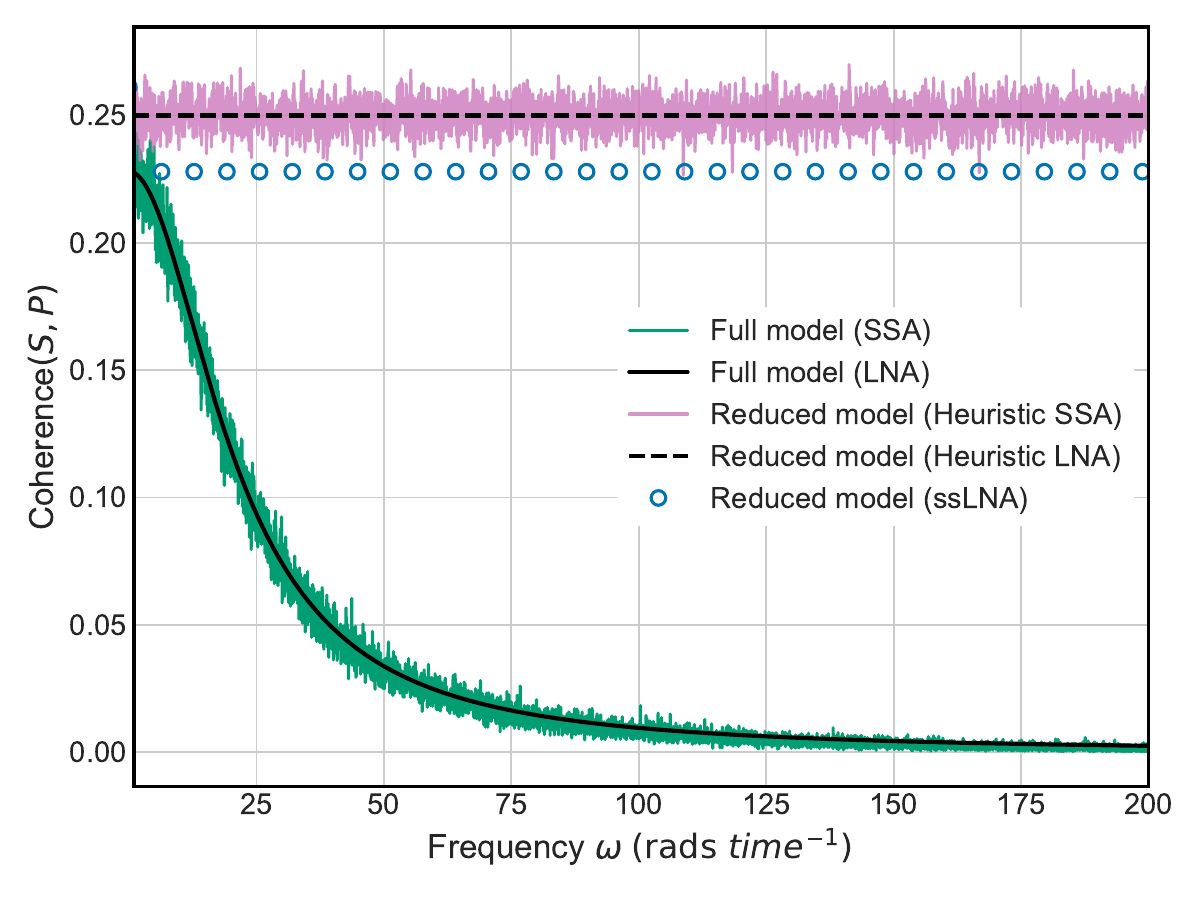}
        \caption{Comparison of the coherence spectrum of the full and reduced models for the enzyme-catalyzed reaction. The parameter set is chosen so that there the timescale separation condition is enforced; see the main text for the corresponding parameter values. The solid and dashed black lines represent the predictions of the full LNA (Eq.~\eqref{cohfullenzyme}) and the heuristic LNA (Eq.~\eqref{cohheuristicenzyme}), respectively; the prediction of the ssLNA (Eq.~\eqref{cohssLNAenzyme}) is indicated by open blue circles. Estimates for the coherence spectra computed from SSA simulations of the full model (Eq.~\eqref{fullenzyme}) and the heuristic reduced model (Eq.~\eqref{redenzyme}), processed with Welch’s method, are shown in green and pink, respectively. The SSA output was sampled every $0.01$ time units for a total duration of $T_{\max}=10^6$ units; the Nyquist frequency is $\omega_{\rm max} \approx 314$.}
        \label{fig:fig_2}
\end{figure}

We test our hypothesis as follows. We fix the parameter set to $k_1 = 1$, $k_2 = 0.1$, $k_3 = 2$, $k_4 = 20$, $k_5 = 1$, and $\phi_T = 1$, which
satisfies timescale separation conditions, since $\tau_S = 10.53$, $\tau_P = 1$, and $\tau_C = 0.043$. The steady-state predictions from the ssLNA and the full LNA for the variances and covariances of the concentration fluctuations of substrate and product are very similar; to six significant figures, these read $C_S^{\text{full}} = 11.6617$, $C_S^{\text{ssLNA}} = 11.6620$; $C_P^{\text{full}} = 0.961920$, $C_P^{\text{ssLNA}} = 0.960092$; and $C_{S,P}^{\text{full}} = 0.0399845$, $C_{S,P}^{\text{ssLNA}} = 0.0399076$, which implies that the relative errors are less than $0.2\%$. However, the ssLNA yields very poor estimates for the coherence spectrum, as seen in Fig.~\ref{fig:fig_2}; in fact, its prediction is similar to the coherence spectrum predicted by the heuristic LNA. We also estimate the coherence spectra directly by applying Welch’s method \cite{welch1967use}, as implemented in the {\tt Scipy.signal} package, to trajectories generated by the Gillespie algorithm \cite{gillespie1977exact} of the full and reduced reaction networks in Eqs.~\eqref{fullenzyme} and \eqref{redenzyme}, respectively. For details on the implementation of the Gillespie algorithm for the reduced model, see \cite{myfootnote,rao2003stochastic}. Since the coherence spectra from simulations agree very well with the predictions of the full LNA and the heuristic LNA, it follows that the differences between these two types of LNA are not related to the continuum approximation that is inherent in the LNA. 


In particular, we note that the coherence spectrum of the full LNA always decays to zero, while both the ssLNA and the heuristic LNA predict a constant coherence in the large-$\omega$ limit --- this discrepancy is evident from Eqs.~\eqref{cohfullenzyme} through \eqref{cohheuristicenzyme}, in fact, which show that the coherence decays as $\omega^{-2}$ except for the special case when $x = 1/2$, where it decays as $\omega^{-4}$; see Appendix~\ref{AppB}. The resulting differences in the tails of the coherence spectra agree with the general derivation in Section \ref{IVB} --- they stem from the fact that there are no reactions in Eq.~\eqref{fullenzyme} which cause a simultaneous change in the net numbers of substrate and product molecules, whereas such a reaction is present in the reduced model, Eq.~\eqref{redenzyme}. The different scaling laws for large $\omega$ predicted by the full and reduced models mirror those in the cross-spectrum, see Appendix~\ref{AppB}. By contrast, note that the differences between the predicted coherences are smallest for low $\omega$, which is also borne out by theory:
\begin{align}
     \xi_{S,P}^{\text{full}}(0) = \xi_{S,P}^{\text{ssLNA}}(0) = \frac{1}{4} \frac{1}{1 + K x (x-1)} \ne \xi_{S,P}^{\text{heur}}(0),
     \label{smallwenzyme}
\end{align}
where $K = k_4/(k_3 + k_4)$. In the example of Figure \ref{fig:fig_2}, $x=1/20$ is very small; therefore, all three predicted coherences are in good agreement when $\omega = 0$. The exactness of the ssLNA prediction at $\omega = 0$ is a special case of the general derivation in Section \ref{IVA}. 

Finally, we seek to understand the underlying reason for the inaccuracy of the coherences predicted by the reduced LNAs. It can be shown that the limit of large $k_3$ in Eqs.~\eqref{cohfullenzyme} and \eqref{cohssLNAenzyme} at fixed $\omega$ results in convergence of the second-order moments of the full LNA and the heuristic LNA: $C_S^{\text{full}}/C_S^{\text{heur}} \rightarrow 1$, $C_P^{\text{full}}/C_P^{\text{heur}} \rightarrow 1$, and $C_{S,P} ^{\text{full}} - C_{S,P}^{\text{heur}} \rightarrow 0$. The same is found for the ssLNA. In addition, the coherences predicted by the three LNAs also converge to the same value:
\begin{align}
     \xi_{S,P}^{\text{full}}(\omega),\ \xi_{S,P}^{\text{ssLNA}}(\omega) \rightarrow \frac{1}{4} = \xi_{S,P}^{\text{heur}}(\omega).
     \label{simplecase}
\end{align}
In contrast, the full LNA, Eq.~\eqref{cohfullenzyme}, predicts that if $x \ne 1/2$, then the coherence spectrum decays as
\begin{align}
     \xi_{S,P}^{\text{full}}(\omega) \sim \frac{1}{4} \frac{(2 x-1)^2}{(x-1)^2} \frac{k_3 k_4}{\omega^2}
     \label{enztails}
\end{align}
for large frequencies in the timescale separation limit of large $k_3$,
whereas for $x = 1/2$, the spectrum decays as
\begin{align}
     \xi_{S,P}^{\text{full}}(\omega) \sim \frac{4 k_3^3 k_4}{\omega^4}.
     \label{enztails1}
\end{align}
The discrepancies between the expressions in Eqs.~\eqref{simplecase}, \eqref{enztails}, and \eqref{enztails1} are due to the timescale separation limit and the large-$\omega$ limit \textit{not commuting}. In Eq.~\eqref{simplecase}, we took the limit of large $k_3$ at fixed $\omega$, while in Eqs.~\eqref{enztails} and \eqref{enztails1}, we first took the limit of large $\omega$ in Eq.~\eqref{cohfullenzyme}, followed by the large-$k_3$ limit. We note that the limit taken at fixed $\omega$ implicitly assumes that the largest (Nyquist) frequency $\omega_{\rm max}$ at which the coherence spectrum can be estimated is always much smaller than the value of $k_3$. By contrast, in stochastic simulations, for any finitely large value of $k_3$ it is possible to measure the coherence spectrum for any $\omega$, since $\omega_{\rm max}$ can be made arbitrarily large by increasing the sampling rate of the timeseries of substrate and product molecule numbers output by the Gillespie algorithm. In a real-world scenario, the Nyquist frequency is set by the sampling rate of the device used to make molecule number measurements; if $k_3 \gg \omega_{\rm max}$, then the large-frequency tail of the measured coherence spectrum would be consistent with Eq.~\eqref{simplecase} in that case, whereas if $k_3 \ll \omega_{\rm max}$, then the tails would be described by Eqs.~\eqref{enztails} and \eqref{enztails1}. In other words, the full LNA can accurately describe both cases, while the reduced LNAs can only approximate the low-Nyquist-frequency case.

Hence, we conclude that the coherence spectra predicted by reduced models of enzyme catalysis are generally not accurate, even if parameter sets are chosen so that molecule numbers are predicted accurately. These results have implications for the MI rate between the substrate and the product which is given by Eq.~\eqref{MIrate}. Since $\xi_{S,P}^{\text{heur}}$ is a constant and $\xi_{S,P}^{\text{ssLNA}}$ tends to a constant for large $\omega$, the MI rate is infinite according to the heuristic LNA and the ssLNA. Since the full LNA predicts $\xi_{S,P}^{\text{full}} \propto\omega^{-2}$ for large $\omega$, it follows from Appendix~\ref{AppA} that the MI rate is always finite and positive, even when timescale separation applies. In fact, the MI rate can be quite small, in contradiction to the infinite rate predicted by the reduced LNAs. For example, for the parameter set used in Fig.~\ref{fig:fig_2}, the MI rate computed by substituting the coherence predicted by the full LNA, Eq.~\eqref{cohfullenzyme}, into Eq.~\eqref{MIrate}, is merely $1.83$ bits per unit time.

\subsection{Models of protein translation}

Next, we consider two different models of protein translation and calculate, for each of those, the coherence and MI rate between mRNA and protein fluctuations. We hence quantify the information flow from mRNA to protein, which is a topic that has attracted significant attention in the past two decades, as the correlation between the two is not as high as expected \cite{maier2009correlation,payne2015utility,buccitelli2020mrnas,popovic2018multivariate}.

\subsubsection{Simple model of translation}

Consider the following system of reactions:
\begin{align}
    \varnothing \xrightarrow{k_1} M \xrightarrow{k_2} \varnothing,\quad M \xrightarrow{k_3}C,\quad C\xrightarrow{k_4}C+P,\quad C\xrightarrow{k_5}M,\quad P\xrightarrow{k_6}\varnothing.
    \label{transmodelfull}
\end{align}
We model translation using a minimal reaction scheme in which cytoplasmic mRNA $M$ is transcribed at rate $k_{1}$ and degraded at rate $k_{2}$. Free mRNA can bind ribosomes, which are assumed to be in large excess, forming a translating complex $C$ at rate $k_{3}$. From complex, protein $P$ is synthesized at rate $k_{4}$ without complex being consumed. Ribosome unbinding from the mRNA occurs at rate $k_{5}$, releasing the mRNA. Finally, proteins are degraded (or diluted) at rate $k_{6}$. We refer to the LNA corresponding to this circuit as the full LNA.

From the deterministic rate equations, one finds that the timescales of the three species are given by
\begin{align}
    \tau_M = \frac{1}{k_2+k_3},\quad\tau_P = \frac{1}{k_6},\quad\text{and}\quad\tau_C = \frac{1}{k_5}. 
\end{align}
When $\tau_C \ll \tau_M, \tau_P$, perturbations about the steady-state concentration of complex decay much faster than those about the steady-state concentrations of mRNA and protein. The complex is then in a quasi-steady state, with the deterministic rate equations simplifying to those of an effective reduced circuit,
\begin{align}
    \varnothing \xrightarrow{k_1} M \xrightarrow{k_2} \varnothing,\quad M \xrightarrow{k'}M + P,\quad P\xrightarrow{k_6}\varnothing,
    \label{transmodelred}
\end{align}
where $k' = k_3 k_4 / k_5$. We refer to the LNA corresponding to this circuit as the heuristic LNA. A second, more rigorous way to obtain a reduced LNA when complex is fast relies on the ssLNA. The corresponding calculations can be found in Appendix~\ref{AppC}.

The predictions obtained from the three LNAs for the steady-state mean concentrations coincide exactly. However, the predicted second moments at steady state differ. The timescale separation condition corresponds to $\tau_M,\tau_P\gg\tau_C$. There are different options for enforcing that condition, with varying consequences for the relationships between the second moments resulting from the three LNAs. For example, if we set $k_2 = z$, $k_3 = a_1 z$, and $k_6= a_2 z$, with $a_1$ and $a_2$ some positive constants, and take the limit of $z \rightarrow 0$, then these relationships simplify to
\begin{align}
    \frac{C_{M}^{\text{full}}}{C_{M}^{\text{heur}}} &= 1, 
    \quad\frac{C_{M,P}^{\text{full}}}{C_{M,P}^{\text{heur}}} = 1, 
    \quad\frac{C_{P}^{\text{full}}}{C_{P}^{\text{heur}}} = 1 + \frac{(a_2 + 1)\lambda}{a_1 \lambda + a_2 + 1}, \\[6pt]
    \frac{C_{M}^{\text{full}}}{C_{M}^{\text{ssLNA}}} &= 1, 
    \quad\frac{C_{M,P}^{\text{full}}}{C_{M,P}^{\text{ssLNA}}} = 1, 
    \quad\text{and}\quad\frac{C_{P}^{\text{full}}}{C_{P}^{\text{ssLNA}}} = 1,
\end{align}
where $\lambda = k_4/k_5$. Note that while the ssLNA perfectly agrees with the full LNA in this limit, the heuristic LNA does not agree with the full LNA unless the further constraint of $\lambda \rightarrow 0$ is enforced, which we shall call the ``low translation constraint". A different way to impose a timescale separation is to take $k_5 \rightarrow \infty$, in which case all three LNAs agree: in this limit, one automatically enforces both the timescale separation condition $\tau_M,\tau_P\gg\tau_C$ and the low translation constraint ($\lambda \rightarrow 0$). 

The coherence spectra of mRNA and protein, as predicted by the full LNA, the heuristic LNA, and the ssLNA are given by
\begin{align}
    \xi_{M,P}^{\text{full}}(\omega) &= \frac{\alpha  \lambda  \big(\gamma -\omega ^2 \tau
   _C^2\big)^2}{\big[\gamma +\omega ^2 (\alpha +\gamma )
   \tau _C^2\big]\big\{\gamma [\lambda (\alpha +\gamma
   )+\gamma ]+\omega ^2 \tau _C^2 \big[\alpha ^2+2 \alpha 
   (\gamma +1)+\gamma ^2+\lambda +1\big]+\omega ^4 \tau
   _C^4\big\}},
    \label{cohfulltrans} \\
    \xi_{M,P}^{\text{ssLNA}}(\omega) &= \frac{\alpha  \gamma  \lambda }{\gamma [\lambda  (\alpha
   +\gamma )+\gamma ]+(\lambda +1) \omega ^2 \tau _C^2},\quad\text{and}
    \label{cohssLNAtrans} \\
    \xi_{M,P}^{\text{heur}}(\omega) &= \frac{\alpha  \gamma  \lambda }{\gamma  (\alpha  \lambda
   +\gamma )+\omega ^2 \tau _C^2}, 
    \label{cohheuristictrans}
\end{align}
respectively, where $\alpha = k_3/k_5$, $\beta = k_6/k_5$, $\lambda = k_4/k_5$, and $\gamma = k_2/k_5$. Note that $\alpha+\gamma = \tau_C/\tau_M$ and $\beta = \tau_C/\tau_P$.

The coherences predicted by the full LNA and the ssLNA agree exactly at $\omega = 0$, while the coherence of the heuristic LNA differs:
\begin{align}
     \xi_{M,P}^{\text{full}}(0) = \xi_{M,P}^{\text{ssLNA}}(0) = \frac{\alpha  \lambda }{\lambda  (\alpha +\gamma )+\gamma } \ne \xi_{M,P}^{\text{heur}}(0) = \frac{\alpha  \lambda }{\alpha  \lambda +\gamma }.
     \label{smallwtrans}
\end{align}
These relationships constitute a special case of the general derivation in Section~\ref{IVA}. We also note that the large-frequency tails of the coherence spectra of the three LNAs scale as $\omega^{-2}$, which follows directly from the general results in Section~\ref{IVB} --- the agreement stems from the fact that in both the full model, Eq.~\eqref{transmodelfull}, and the reduced model, Eq.~\eqref{transmodelred}, no reactions occur that cause a simultaneous change in the net numbers of mRNA and product molecules. 

However, we note that for intermediate frequencies, there is a major difference between the coherence spectra. While both the ssLNA and the heuristic LNA predict a monotonically decreasing coherence spectrum with increasing $\omega$, the full LNA does not. Rather, it predicts that as $\omega$ increases from $0$, the coherence first decreases and attains the value of $0$ at the critical frequency $\omega_c = \sqrt{\gamma}/\tau_C = \sqrt{k_2 k_5}$; it then increases again, reaching a peak before decreasing as $\omega^{-2}$. Note that this non-monotonic behavior is present for all finite values of the rate constants, and that it is hence generally impossible to recover the monotonicity predicted by the reduced LNAs.

\begin{figure}[h]
        \centering       
        \includegraphics[scale=0.65]{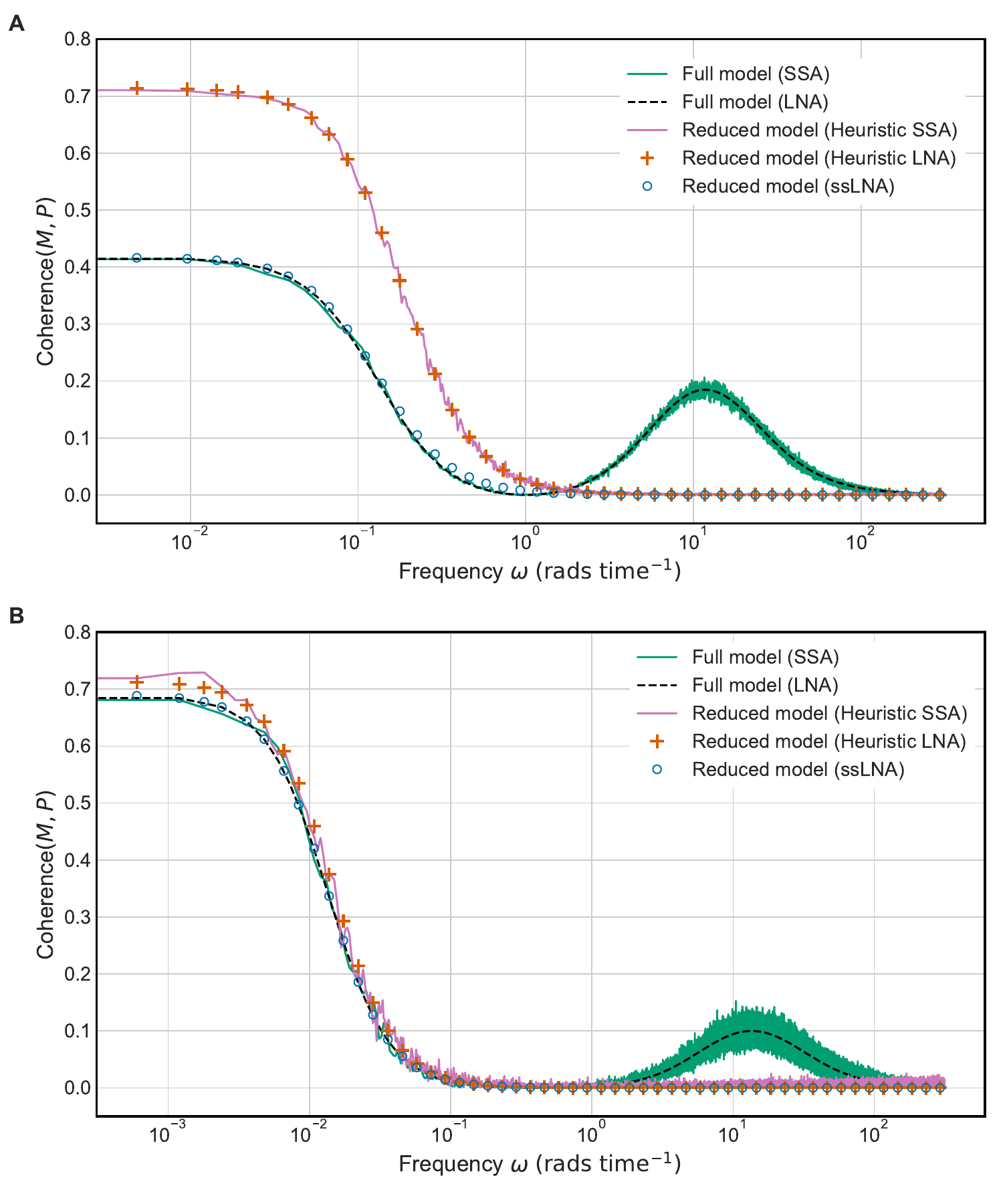}
        \caption{Comparison of the coherence spectrum of the full and reduced models for the simple translation model. We consider two parameter sets: (A) $k_1 = 10$, $k_2 = 0.1$, $k_3 = 0.1$, $k_4 = 25$, $k_5 = 10$, and $k_6 = 0.05$; (B) $k_1 = 10$, $k_2 = 0.0075$, $k_3 = 0.1$, $k_4 = 5$, $k_5 = 25$, and $k_6 = 0.005$. Both enforce timescale separation conditions, with $\tau_C/\tau_M = 0.02$ and $\tau_C/\tau_P = 0.005$ for (A), and $\tau_C/\tau_M = 0.0043$ and $\tau_C/\tau_P = 0.0002$ for (B). While the first parameter set does not obey the low translation constraint, the second set does, since $k_4 \ll k_5$. The coherence was estimated numerically with Welch's method from SSA simulations of the full model (Eq.~\eqref{transmodelfull}) and the reduced model (Eq.~\eqref{transmodelred}); the SSA output was sampled every $0.01$ time units for a total duration of a few million time units. The Nyquist frequency is $\omega_{\rm max} \approx 314$. We also show the coherence predicted by the theory in Eqs.~\eqref{cohfulltrans} through \eqref{cohheuristictrans} for the full LNA, the ssLNA, and the heuristic LNA. 
        }
        \label{fig:fig_3}
\end{figure}

As an example, we reconsider one of the limits applied previously to obtain a timescale separation: setting $k_2 = z$, $k_3 = a_1 z$, and $k_6= a_2 z$, where $a_1$ and $a_2$ are some positive constants, and taking the limit of $z \rightarrow 0$ at fixed $\omega$, we find $w_c \rightarrow 0$, which means that the coherence spectrum of the full LNA now exhibits a coherence of $0$ at $\omega = 0$, increases to reach a peak, and then decreases as $\omega^{-2}$ which still does not agree with the monotonically decreasing behavior predicted by the ssLNA and the heuristic LNA. (Still, in this limit, the first and second moments of mRNA and protein noise predicted by the ssLNA agree exactly with those of the full LNA.) We can also consider the second limit introduced before to obtain a timescale separation, with $k_5 \rightarrow \infty$ at fixed $\omega$. We first note that the latter is impossible to achieve practically, as any chosen value of $k_5$ will still be finite despite being very large --- hence, $\omega_c$ is large, but finite, which implies that if the SSA output is sampled on a sufficiently fine time grid, then the coherence spectrum will always display a peak before decaying as $\omega^{-2}$, in contradiction to the monotonic predictions of the ssLNA and the heuristic LNA. 

The non-vanishing differences between the full LNA and the reduced LNAs are also apparent if we consider the large-frequency tails of the coherence spectra:
\begin{align}
     \xi_{M,P}^{\text{full}}(\omega)& \sim \frac{\alpha  \lambda }{(\alpha +\gamma ) \tau _C^2} \frac{1}{\omega^2}, \\ \xi_{M,P}^{\text{ssLNA}}(\omega)& \sim \frac{\alpha  \gamma  \lambda }{(\lambda +1) \tau _C^2} \frac{1}{\omega^2},\quad\text{and} \\ \xi_{M,P}^{\text{heur}}(\omega)& \sim \frac{\alpha  \gamma  \lambda }{\tau _C^2} \frac{1}{\omega^2}.
\end{align}
If we again impose the low translation constraint ($\lambda\to 0)$, then the ssLNA and the heuristic LNA agree; however, the full LNA only agrees with those if we apply the additional constraint that $\gamma = (\alpha + \gamma)^{-1}$. That constraint is impossible to satisfy under timescale separation conditions, however, since it requires $\alpha$ and $\gamma$ to be very small due to $\alpha + \gamma = \tau_C/\tau_M$.  

We tested some of these analytic conclusions by means of stochastic simulations; the results are shown in Fig.~\ref{fig:fig_3}A. We chose a parameter set that enforces timescale separation conditions, with $\tau_C/\tau_M = 0.02$ and $\tau_C/\tau_P = 0.005$, and then compared the coherence spectra predicted by the full LNA, the ssLNA, the heuristic LNA, the SSA of the full model, Eq.~\eqref{transmodelfull}, and the SSA of the reduced model, Eq.~\eqref{transmodelred}. As expected, the full LNA and the SSA of the full model agreed exactly, as did the heuristic LNA and the SSA of the reduced model. The coherence at $\omega = 0$ of the full LNA was perfectly predicted by the ssLNA, which accurately predicts the entire LNA spectrum up to $\omega \approx \omega_c = 1$, but becomes inaccurate for higher frequencies -- in particular, since the ssLNA prediction is monotonically decreasing with frequency, it completely misses the peak at $\omega \approx 10$. Note that the second moments of the full LNA and the ssLNA are very close, as claimed above: $\frac{(\sigma_M^2)^{\text{full}}}{(\sigma_M^2)^{\text{ssLNA}}} = 1$, $\frac{(\sigma_{MP})^{\text{full}}}{(\sigma_{MP})^{\text{ssLNA}}} = 0.992$, and $\frac{(\sigma_P^2)^{\text{full}}}{(\sigma_P^2)^{\text{ssLNA}}} = 0.993$. By contrast, the heuristic LNA prediction is inaccurate for all values of $\omega$, which stems from the fact that the chosen parameter set enforces timescale separation conditions, but not the additional low translation constraint of small $\lambda$. When this additional constraint is applied, the ssLNA and the heuristic LNA are of similar accuracy (Fig.~\ref{fig:fig_3}B). 

These results have significant implications for the MI rate. For the example in Fig.~\ref{fig:fig_3}B, we apply Eq.~\eqref{MIrate} to compute the MI rate in bits per unit time using the full LNA, the ssLNA, and the heuristic LNA:
\begin{align}
    M_{M,P}^{\text{ full}} = 1.26,\quad M_{M,P}^{\text{ ssLNA}} = 4.3 \times 10^{-3},\quad\text{and}\quad M_{M,P}^{\text{ heur}} = 4.9 \times 10^{-3}.
\end{align}
Hence, the inaccuracies in the coherence spectrum predicted by the reduced LNAs for intermediate and large frequencies lead to a massive underestimation, by a factor of about $300$, of the MI rate.

\subsubsection{Model of translation with compartmentalization and transcriptional bursting}\label{trans2}

Consider the system of reactions
\begin{align}
    \varnothing \xrightarrow{k_1} r M_n \xrightarrow{k_2} \varnothing,\quad M_n \xrightarrow{k_3}M_c,\quad M_c\xrightarrow{k_4}\varnothing,\quad M_c \xrightarrow{k_5}M_c + P,\quad P\xrightarrow{k_6}\varnothing,
    \label{transmodelfull1}
\end{align}
in which basal transcription produces nuclear mRNA ($M_{n}$) from DNA in bursts of size $r$. We will assume that $r$ is drawn from a geometric distribution with mean $b$ (the mean burst size) \cite{jia2017simplification,golding2005real}, which models transcriptional bursting. Newly synthesized $M_{n}$ can be degraded in the nucleus or exported to the cytoplasm, where it is labelled as cytoplasmic mRNA ($M_{c}$). Cytoplasmic $M_{c}$ is translated by ribosomes to produce protein $P$, with both $M_{c}$ and $P$ subject to first-order degradation. For simplicity, we do not explicitly model the mRNA-ribosome complex $C$, as we did in the reaction scheme in Eq.~\eqref{transmodelfull}; however, qualitatively similar results follow if we add that additional species. In what follows, we refer to the LNA corresponding to the circuit in Eq.~\eqref{transmodelfull1} as the full LNA.

From the deterministic rate equations, one finds that the timescales of the three species are given by
\begin{align}
    \tau_{M_n} = \frac{1}{k_2+k_3},\quad \tau_{M_c} = \frac{1}{k_4},\quad\text{and}\quad\tau_P = \frac{1}{k_6}. 
    \label{tsconds1}
\end{align}
When $\tau_{M_n} \ll \tau_{M_c}, \tau_P$, perturbations about the steady-state concentration of nuclear mRNA decay much faster than those about the steady-state concentrations of cytoplasmic mRNA and protein. Nuclear mRNA is then in a quasi-steady state, with the deterministic rate equations simplifying to those of an effective reduced circuit,
\begin{align}
    \varnothing \xrightarrow{k'} r M_C \xrightarrow{k_4} \varnothing,~~~M_C \xrightarrow{k_5}M + P,~~~P\xrightarrow{k_6}\varnothing,
    \label{transmodelred1}
\end{align}
where $k' = k_1 k_3/(k_2 + k_3)$. The properties of the random variable $r$ are as for the full circuit. We refer to the LNA corresponding to Eq.~\eqref{transmodelred1} as the heuristic LNA. A second, more rigorous way to obtain a reduced LNA when complex is fast relies on the ssLNA. The corresponding calculations can be found in Appendix~\ref{AppD}.

In the limit of $k_2 \rightarrow \infty$, nuclear mRNA is very short-lived compared to the other two species, see Eq.~\eqref{tsconds1}, and the second moments at steady state predicted by the ssLNA match those in the full LNA; however, those predicted by the heuristic LNA do not:
\begin{align}
    \frac{C_{M_c}^{\text{full}}}{C_{M_c}^{\text{heur}}} &= \frac{1}{1 + b}, 
    \quad\frac{C_{{M_c},P}^{\text{full}}}{C_{{M_c},P}^{\text{heur}}} = \frac{1}{1 + b}, 
    \quad\frac{C_{P}^{\text{full}}}{C_{P}^{\text{heur}}} = \frac{k_4 + k_5 + k_6}{k_4 + k_5(1 + b) + k_6}, \notag \\[6pt]
    \frac{C_{M_c}^{\text{full}}}{C_{M_c}^{\text{ssLNA}}} &= 1, 
    \quad\frac{C_{M_c,P}^{\text{full}}}{C_{M_c,P}^{\text{ssLNA}}} = 1, 
    \quad\text{and}\quad\frac{C_{P}^{\text{full}}}{C_{P}^{\text{ssLNA}}} = 1. \label{vareqs}
\end{align}
 For the heuristic LNA to agree with the full LNA and the ssLNA, the mean burst size $b$ must be much smaller than $1$; that is an unrealistic constraint, since typical burst sizes for genes in eukaryotic cells are approximately in the range of $2-140$, see Table I of \cite{cao2020analytical}. 
\begin{figure}[h]
        \centering       
        \includegraphics[scale=0.7]{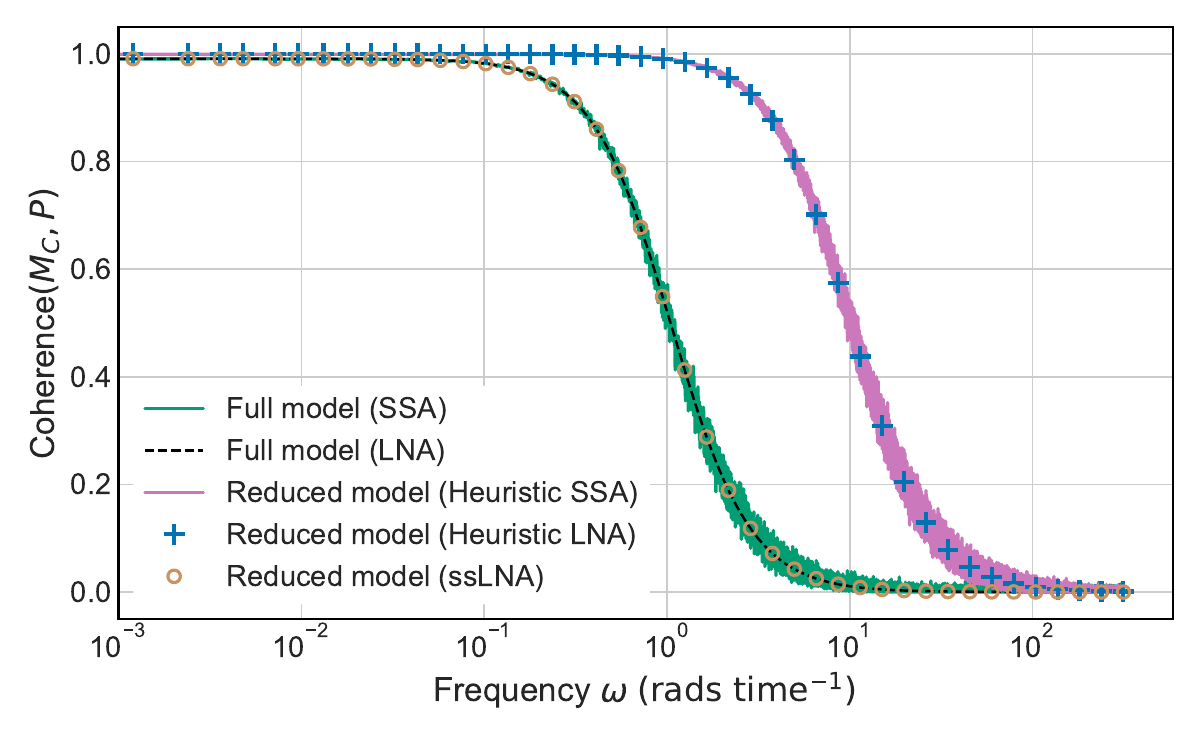}
        \caption{Comparison of the coherence spectrum of the full and reduced models for a model of translation that includes transcriptional bursting and nuclear export. We choose the parameter set $b = 100, k_1 = 5$, $k_2 = 1000$, $k_3 = 1$, $k_4 = 0.1$, $k_5 = 10$, and $k_6 = 0.05$, which enforces timescale separation conditions due to $\tau_{M_n}/\tau_{M_c} = 10^{-4}$ and $\tau_{M_n}/\tau_P = 5 \times 10^{-5}$. The coherence was estimated numerically with Welch's method from SSA simulations of the full model (Eq.~\eqref{transmodelfull1}) and the reduced model (Eq.~\eqref{transmodelred1}); the SSA output was sampled every $0.01$ time units for a total duration of a few million time units. The Nyquist frequency is $\omega_{\rm max} \approx 314$. We also show the coherence predicted by the theory in Eqs.~\eqref{cohfulltrans1} through \eqref{cohheuristictrans1} for the full LNA, the ssLNA, and the heuristic LNA.}
        \label{fig:fig_4}
\end{figure} 

The coherence spectra of cytoplasmic mRNA and protein predicted by the full LNA, the heuristic LNA, and the ssLNA are given by
\begin{align}
    \xi_{M_c,P}^{\text{full}}(\omega) &= \frac{k_4 k_5 \big\{(k_2 + k_3)[k_2 + k_3 (1 + b)] + \omega^2 \big\}}
{(k_2 + k_3) k_4 \big\{ k_2 (k_4 + k_5) + k_3 [k_4 + k_5 (1 + b)] \big\} 
+ \big[ (k_2 + k_3)^2 + k_4 (k_4 + k_5) \big] \omega^2 + \omega^4},
    \label{cohfulltrans1} \\
    \xi_{M_c,P}^{\text{ssLNA}}(\omega) &= \frac{[k_2 + k_3 (1 + b)]\, k_4 k_5}
{k_2 k_4 (k_4 + k_5) + k_3 k_4 [k_4 + k_5 (1 + b)] + (k_2 + k_3)\,\omega^2},\quad\text{and}
    \label{cohssLNAtrans1} \\
    \xi_{M_c,P}^{\text{heur}}(\omega) &= \frac{(1 + b)\,k_4 k_5}{k_4 [k_4 + k_5 (1 + b)] + \omega^2},
    \label{cohheuristictrans1}
\end{align}
respectively. As expected, in the limit of zero frequency, the full LNA and the ssLNA agree exactly, but the heuristic LNA does not. In the limit of large frequencies, the coherence spectra reduce to
\begin{align}
     \xi_{M_c,P}^{\text{full}}(\omega)& \sim  k_4 k_5 \frac{1}{\omega^2}, \label{cohtail1} \\ \xi_{M_c,P}^{\text{ssLNA}}(\omega)& \sim \frac{(k_2 + k_3 + b k_3)\,k_4 k_5}{k_2 + k_3} \frac{1}{\omega^2},\quad\text{and} \label{cohtail2} \\ \xi_{M_c,P}^{\text{heur}}(\omega)& \sim (1 + b) k_4 k_5 \frac{1}{\omega^2}. \label{cohtail3}
\end{align}
Clearly, in the limit of timescale separation, with $k_2 \rightarrow \infty$, the tails of the coherence spectra of the ssLNA and the full LNA agree, which is the first example in this paper where such agreement is found: for the metabolic and other translation models that we previously studied, the tails of the coherence spectra did not match in the timescale separation limit. 

In fact, we find that in this limit, it is not merely the tails, but also the full coherence spectra of the full LNA and the ssLNA which converge to the same expression,
\begin{align}
     \xi_{M_c,P}^{\text{full}}(\omega),\ \xi_{M_c,P}^{\text{ssLNA}}(\omega) \rightarrow \frac{k_4 k_5}{k_4 (k_4 + k_5) + \omega^2}, 
\end{align}
while the coherence spectrum of the heuristic LNA converges to a different limit:
\begin{align}
    \xi_{M_c,P}^{\text{heur}}(\omega) \rightarrow \frac{(1 + b) k_4 k_5}{k_4 [k_4 + k_5 (1 + b)] + \omega^2}.
\end{align}
Correspondingly, the MI rates computed from Eq.~\eqref{MIrate} are given by
\begin{align}
  M_{M_c,P}^{\text{full}},\ M_{M_c,P}^{\text{ssLNA}} &\rightarrow  \frac{\sqrt{k_4 (k_4 + k_5)} - k_4}{\ln 4}\quad\text{and} \\
  M_{M_c,P}^{\text{heur}}& \rightarrow  \frac{\sqrt{k_4 [k_4 + (1 + b) k_5]} - k_4}{\ln 4}.
\end{align}
Hence, we have shown that it is possible for the coherence spectrum and the MI rate to be correctly preserved when a model is reduced using the ssLNA. The underlying reason is likely that here, the fast species (nuclear mRNA) affects the slow species (cytoplasmic mRNA and protein), but that it itself is not affected by them, in contrast to previous models. However, our results also show that use of the commonly applied heuristic method of model reduction does not give the correct coherence spectrum and MI rate in the limit of timescale separation; in particular, the MI rate is always overestimated by the heuristic method. In Fig.~\ref{fig:fig_4}, we verify our analytical results for the coherence spectra using stochastic simulation.

Note that in the aforementioned discussion, we have considered the limit of timescale separation by taking the limit as $k_2 \rightarrow \infty$. A different way to enforce the timescale separation condition $(\tau_{M_n} \ll \tau_{M_c}, \tau_P)$ involves taking the limit of $k_3 \rightarrow \infty$, see Eq.~\eqref{tsconds1}, which leads to different results than previously reported: (i) the second moments of the full LNA, the ssLNA, and the heuristic LNA agree exactly in this limit, recall~Eq.~\eqref{vareqs}); and (ii) the tails of the coherence spectra of the full LNA and the ssLNA do not agree in the timescale separation limit. The latter can be directly deduced by inspection of Eqs.~\eqref{cohtail1} through \eqref{cohtail3} which show that in the limit as $k_3 \rightarrow \infty$, the coherence spectra predicted by the heuristic LNA and the ssLNA converge to a single common expression that is different from the one obtained from the full LNA. Hence, we have shown that it is not simply the method of model reduction which affects these predictions, but also \textit{the particular parameter limit by which timescale separation is enforced}. 

\section{Discussion}

In this paper, we have demonstrated that although model reduction methods may reliably estimate the molecule number statistics of the underlying full models under timescale separation conditions, they do not necessarily provide accurate estimates of coherence spectra or MI rates. We have shown that, even when reduced models are derived using rigorous methods such as the ssLNA, they can substantially misestimate coherence, particularly at intermediate and high frequencies, consequently leading to either significantly lower or significantly higher values for the MI rates than are predicted by the corresponding full models. Moreover, we have shown that commonly used heuristic reduction methods can further exacerbate these inaccuracies by neglecting noise from fast reactions. 

One of the main issues we identified is that reduced models may incorrectly predict the scaling laws characterizing the tails of coherence spectra, with considerable repercussions for the calculation of the associated MI rates. That issue inherently results from the fact that model reduction implicitly involves a timescale separation limit, with one or more parameter values in the full model tending to either zero or infinity at \textit{constant} frequency, a procedure that generally does not faithfully preserve the large-frequency tails of coherence spectra. For example, in the enzyme-catalyzed system, Eq.~\eqref{fullenzyme}, the LNA predicts that for all parameter values, including those enforcing timescale separation, the coherence scales as $\omega^{-2}$ for large frequencies, whereas the LNA of the reduced model, Eq.~\eqref{redenzyme}, predicts a constant that is independent of the frequency and the parameter set. Another example is the simple protein translation model, Eq.~\eqref{transmodelfull}, and its reduced version, Eq.~\eqref{transmodelred}, where the tails of the coherence spectra scale as $\omega^{-2}$, but with different prefactors that do not converge to each other in the timescale separation limit. In that same example, the full model predicts that for all parameters, there exists a peak in the coherence spectrum at intermediate frequencies, whereas the reduced models predict a monotonically decreasing coherence spectrum.    

Furthermore, we have shown that, while model reduction may not generally preserve coherence spectra and MI rates, there are specific cases where it does; for example, when modeling compartmentalization, transcriptional bursting, and protein translation in Section \ref{trans2}, the full model and the rigorously reduced model from the ssLNA predicted the same coherence spectra for \textit{all frequencies} when timescale separation was enforced by taking the nuclear mRNA degradation rate to infinity. The precise conditions under which such perfect agreement arises remain unclear; however, it is likely in part attributable to the absence of feedback from the slow species to the fast species in the full model. That cannot be the only reason, though, as we have also shown that if timescale separation is enforced differently, by taking the nuclear export rate to infinity, the coherence spectrum of the ssLNA no longer converges to that of the full LNA. Hence, our results indicate that the coherence spectrum and the MI rate are in general not uniquely defined under timescale separation conditions.  

We note that constant coherence at large frequencies, and the associated infinite MI rate, were previously identified for the simple biochemical system $\emptyset \rightarrow A \rightarrow \emptyset$, $A \rightarrow B \rightarrow \emptyset$; see Motif II in \cite{tostevin2009mutual,tostevin2010mutual} and Motif b in \cite{moor2025state}. Our results in Section~\ref{IVB} extend this special case: we show that any biochemical system to which the LNA is applicable and which contains at least one chemical reaction that simultaneously changes the copy numbers of species $A$ and $B$ yields constant high-frequency coherence between these species under the LNA or its reduced forms. In \cite{moor2025state}, it was further shown that for the circuit mentioned above, a finite MI rate can be recovered if an alternative formulation of the LNA is used which retains reaction-specific information in the trajectories and which does not rely on coherence as an intermediate step in the MI rate calculation. However, from the model reduction point of view in the present paper, a more fundamental issue is that the circuit $\emptyset \rightarrow A \rightarrow \emptyset$, $A \rightarrow B \rightarrow \emptyset$ is simply unrealistic: in reality, it represents the reduced form of a more complex circuit such as $\emptyset \rightarrow A \rightarrow \emptyset$, $A \rightarrow C_1 \rightarrow C_2 \rightarrow \cdots \rightarrow C_N \rightarrow B \rightarrow \emptyset$, where species $C_i$ ($i=1,\dots,N$) are fast intermediates. By the results of Section \ref{IVB}, for such a circuit the coherence decreases as $\omega^{-2}$ in the limit of large frequencies, with finite MI rate, since there is no reaction in which the molecule numbers of $A$ and $B$ change simultaneously when it occurs. In the future, it would be interesting to understand how the calculation of the MI rate via the alternative formulation of the LNA in \cite{moor2025state} is affected by model reduction. Based on the example discussed above, our expectation is that the discrepancies between the MI rates predicted by the full LNA and the ssLNA are reduced in comparison to those observed here. 

In conclusion, our results highlight the need for caution when employing reduced models to infer coherence or information-theoretic properties. Given that all models are approximations of reality, the question arises of how to design reduced descriptions that accurately capture not only molecular number statistics, but also the coherences and MI rates of the underlying systems, which can be achieved by constructing systematically models of increasing complexity until further refinements yield negligible changes in their predictions.

\section*{Acknowledgments}
J. D. M. acknowledges support from a PhD scholarship provided by the Darwin Trust. R. G. acknowledges support from the Leverhulme Trust (RPG-2024-082). The authors thank Peter Swain for valuable discussions.

\appendix

\section{Derivation of block-inverse formula for Jacobian of the full model}\label{Appinv}

First, we write the (unknown) inverse of $\underline{J}$ in block form,
\[
\underline{J}^{-1}=\begin{pmatrix} \underline{X} & \underline{Y} \\[4pt] \underline{Z} & \underline{W} \end{pmatrix},
\]
where the blocks $\underline{X}$, $\underline{Y}$, $\underline{Z}$, and $\underline{W}$ have dimensions $N_s\times N_s$, $N_s\times N_f$, $N_f\times N_s$, and $N_f\times N_f$, respectively. Note that from Section~\ref{rigorousred}, we have
\[
\underline{J}=\begin{pmatrix} \underline{J}_{ss} & \underline{J}_{sf} \\[4pt] \underline{J}_{fs} & \underline{J}_{ff} \end{pmatrix}.
\]

The matrix equation $\underline{J}\, \underline{J}^{-1}=\underline{I}_M$,  where $\underline{I}_M$ is the identity matrix of rank $M$, yields the block equations
\begin{align}
\underline{J}_{ss}X + \underline{J}_{sf}Z &= \underline{I}_{N_s}, \label{bl1} \\
\underline{J}_{ss}Y + \underline{J}_{sf}W &= 0, \label{bl2} \\
\underline{J}_{fs}X + \underline{J}_{ff}Z &= 0, \label{bl3} \\
\underline{J}_{fs}Y + \underline{J}_{ff}W &= \underline{I}_{N_f}. \label{bl4}
\end{align}
Assuming that $\underline{J}_{ff}$ is invertible, we solve Eq.~\eqref{bl3} for $Z$ to find
\begin{align}
\underline{Z} = -\,\underline{J}_{ff}^{-1} \underline{J}_{fs} \underline{X}. \label{bl5}
\end{align}
Substituting Eq.~\eqref{bl5} into Eq.~\eqref{bl1}, we can solve for $\underline{X}$:
\begin{align}
\underline{X} = \underline{J}_S^{-1}, \label{bl6}
\end{align}
where we assumed that $\underline{J}_S$, defined in Eq.~\eqref{JSdef}, is invertible. Substitution of Eq.~\eqref{bl6} into Eq.~\eqref{bl5} yields
\begin{align}
\underline{Z} = -\,\underline{J}_{ff}^{-1} \underline{J}_{fs} \underline{J}_S^{-1}. \label{bl7}
\end{align}
Next, we solve for $\underline{Y}$ and $\underline{W}$: from Eq.~\eqref{bl4}, we have
\begin{align}
\underline{W} = \underline{J}_{ff}^{-1}\big(\underline{I}_{N_f} - \underline{J}_{fs}\underline{Y}\big). \label{bl8}
\end{align}
Substituting Eq.~\eqref{bl8} into Eq.~\eqref{bl2} and solving for $\underline{Y}$, we find
\begin{align}
\underline{Y} = -\,\underline{J}_S^{-1} \underline{J}_{sf}\underline{J}_{ff}^{-1}. \label{bl9}
\end{align}
Finally, we substitute Eq.~\eqref{bl9} into Eq.~\eqref{bl8}, which gives the following expression for $\underline{W}$:
\[
\underline{W} = \underline{J}_{ff}^{-1} + \underline{J}_{ff}^{-1} \underline{J}_{fs} \underline{J}_S^{-1} \underline{J}_{sf} \underline{J}_{ff}^{-1}.
\]
Thus, the block-inverse formula for the Jacobian $\underline{J}$ of the full model is
\[
\underline{J}^{-1} =
\begin{pmatrix}
\underline{J}_S^{-1} & -\,\underline{J}_S^{-1} \underline{J}_{sf}\underline{J}_{ff}^{-1} \\[6pt]
-\,\underline{J}_{ff}^{-1}\underline{J}_{fs} \underline{J}_S^{-1} & \; \underline{J}_{ff}^{-1} + \underline{J}_{ff}^{-1} \underline{J}_{fs} \underline{J}_S^{-1} \underline{J}_{sf} \underline{J}_{ff}^{-1}
\end{pmatrix}.
\]

\section{Proof of finiteness of MI rate when $D_{ij}=0$}\label{AppA}

We start by summarizing the well-known properties of the coherence $\xi_{ij}$ between two signals which, in our case, are the real-valued timeseries of the fluctuations of the numbers of molecules of the two species $X_i$ and $X_j$: (i) 
$0 \le \xi_{ij} \le 1$; and (ii) $\xi_{ij}(\omega) = \xi_{ij}(-\omega)$. 

Hence, it follows by Eq.~\eqref{MIrate} that the MI rate can be written as
\begin{align}
    M_{ij} = -\frac{1}{4 \pi \ln 2} \int_{-\infty}^{\infty} \text{ln}(1 - \xi_{ij}(\omega)) d\omega = -\frac{1}{2 \pi \ln 2} \int_{0}^{\infty} \text{ln}(1 - \xi_{ij}(\omega)) d\omega.
    \label{MIrate1}
\end{align}
We divide the integral in Eq.~\eqref{MIrate1} into two parts:
\begin{align}
    M_{ij} = -\frac{1}{2 \pi \ln 2} \biggl(\int_{0}^{R} \text{ln}(1 - \xi_{ij}(\omega)) d\omega + \int_{R}^{\infty} \text{ln}(1 - \xi_{ij}(\omega)) d\omega \biggr),
    \label{MIrate2}
\end{align}
where $R$ is some positive number. The integral from $0$ to $R$ in Eq.~\eqref{MIrate2} is positive and finite, since $\xi_{ij}(\omega)$ is a continuous function of $\omega$ that is bounded between $0$ and $1$. (In particular, $\xi_{ij}(\omega)$ is strictly below $1$, as we are considering open systems in which fluctuations between any two different species can never be perfectly correlated due to synthesis occurring at random points in time.) If $D_{ij} = 0$, then we have proven in Section~\ref{IVB} that $\xi_{ij}(\omega) \propto A w^{-2}$ as $\omega \rightarrow \infty$, where $A$ is a positive number; see Eq.~\eqref{scaling}. We choose $R$ to be sufficiently large so that this scaling law applies. It then follows that  
\[
    \int_{R}^{\infty} \text{ln}(1 - \xi_{ij}(\omega)) d\omega \sim A \int_{R}^{\infty} \omega^{-2} d\omega = A/R > 0.
\]
Hence, it follows from Eq.~\eqref{MIrate2} that if $D_{ij} = 0$, then the MI rate is a finite positive quantity. 

\section{Calculations for enzyme reaction networks} \label{AppB}

\subsection{Full LNA and heuristic LNA}\label{enzymeLNA}

We make use of the LNA frameworks described in Sections~\ref{LNA_MI} and \ref{heurmodelred} to calculate the coherence spectrum for the numbers of substrate and product molecules in the full and reduced enzyme systems under steady-state conditions.

The stoichiometric matrix and rate function vector for the full system, Eq.~\eqref{fullenzyme}, are defined by
\[
\underline{S}_{\text{full}} = \begin{bmatrix}
1 & -1 & 1 & 0 & 0 \\
0 & 0 & 0 & 1 & -1 \\
0 & 1 & -1 & -1 & 0
\end{bmatrix}
\quad\text{and}\quad
\vec{f}_{\text{full}} = \begin{bmatrix}
k_1, & k_2 (\phi_T - \phi_C)\phi_S, & k_3 \phi_C, & k_4 \phi_C, & k_5 \phi_P
\end{bmatrix}^T,
\]
where the first, second, and third rows of $\underline{S}_{\text{full}}$ account for the net changes in the numbers of molecules of species $S$, $P$, and $C$, respectively. Note that free enyzme $E$ does not need to be explicitly accounted for due to the conservation law between free enzyme and complex. 

Similarly, for the heuristically reduced system, Eq.~\eqref{redenzyme}, we can write
\[
\underline{S}_{\text{heur}} = \begin{bmatrix}
1 & -1 & 0 \\
0 & 1 & -1 \\
\end{bmatrix}
\quad\text{and}\quad
\vec{f}_{\text{heur}} = \begin{bmatrix}
k_1, & \frac{k_4 \phi_T \phi_S}{K_M + \phi_S}, & k_5 \phi_P
\end{bmatrix}^T,
\]
where the first and second rows of the stoichiometric matrix $\underline{S}_{\text{heur}}$ account for the net changes in the numbers of molecules of species $S$ and $P$, respectively.

The rate equations for the full system are given by Eq.~\eqref{REq}, with $\underline{S}$ and $\vec{f}$ replaced by $\underline{S}_{\text{full}}$ and $\vec{f}_{\text{full}}$, respectively:
\begin{align}
    \frac{d \phi_S}{d t} &= k_1 - k_2 \phi_S (\phi_T - \phi_C) + k_3 \phi_C, \label{full1} \\
    \frac{d \phi_C}{d t} &= k_2 \phi_S (\phi_T - \phi_C) - (k_3 + k_4) \phi_C, \label{full2}  \\
    \frac{d \phi_P}{d t} &= k_4 \phi_C - k_5 \phi_P.
    \label{full3}
\end{align}
Their solution at steady state leads to the mean concentrations shown in Eq.~\eqref{solenzy}. 

The timescale of species $\phi_i$ is derived via the substitution $\phi_i \mapsto \phi_i + \epsilon_i(t)$ in the rate equations: assuming steady state and solving the resulting differential equation for $\epsilon_i(t)$, we aim to understand how a perturbation in the concentration of species $\phi_i$ about its steady state evolves. Since $\epsilon_i(t) = \epsilon_i(0)e^{-t/\tau_i}$, the perturbation is found to decay exponentially in time. The resulting expressions for the timescales $\tau_i$ $(i = S, P, C)$ are given in Eq.~\eqref{tscond}.

The rate equations for the reduced system, which are obtained from Eq.~\eqref{REq} with $\underline{S}$ and $\vec{f}$ replaced by $\underline{S}_{\text{heur}}$ and $\vec{f}_{\text{heur}}$, respectively, read 
\begin{align}
    \frac{d \phi_S}{d t} &= k_1 -\frac{k_4 \phi_T \phi_S}{K_M + \phi_S}, \label {red1} \\
    \frac{d \phi_P}{d t} &= \frac{k_4 \phi_T \phi_S}{K_M + \phi_S} - k_5 \phi_P.
    \label{red2}
\end{align}
Their solution at steady state is equally given by Eq.~\eqref{solenzy}. Note that the reduced system of rate equations, Eqs.~\eqref{red1} and \eqref{red2}, can be obtained from the rate equations for the full system by setting $d \phi_C/d t= 0$ in Eq.~\eqref{full2}, solving for $\phi_C$, and then substituting the resulting expression into Eqs.~\eqref{full1} and \eqref{full3}. 

Given Eqs.~\eqref{Jac} and \eqref{dmatrix}, the corresponding Jacobian and diffusion matrices are given by
\[
\underline{J}_{\text{full}} =
\begin{bmatrix}
- k_2 \left( \phi_T - \dfrac{k_1}{k_4} \right) & 0 & \dfrac{(k_1 + \phi_T\, k_3) k_4}{\phi_T\, k_4-k_1 } \\[2mm]
0 & -k_5 & k_4 \\[1mm]
k_2 \left(\phi_T - \dfrac{k_1}{k_4}\right) & 0 & -\dfrac{\phi_T\, k_4 (k_3 + k_4)}{\phi_T\, k_4-k_1}
\end{bmatrix}
\quad\text{and}\quad
\underline{J}_{\text{heur}} =
\begin{bmatrix}
-\dfrac{(k_1 - \phi_T k_4)^2}{\phi_T k_4 K_M} & 0 \\
\dfrac{ (k_1 - \phi_T k_4)^2}{\phi_T k_4 K_M} & -k_5
\end{bmatrix},
\]
as well as by
\[
\underline{D}_{\text{full}} =
\begin{bmatrix}
\dfrac{2 k_1 (k_3 + k_4)}{k_4} & 0 & -\dfrac{k_1 (2 k_3 + k_4)}{k_4} \\[1mm]
0 & 2 k_1 & -k_1 \\[1mm]
-\dfrac{k_1 (2 k_3 + k_4)}{k_4} & -k_1 & \dfrac{2 k_1 (k_3 + k_4)}{k_4}
\end{bmatrix}
\quad\text{and}\quad
\underline{D}_{\text{heur}} =
\begin{bmatrix}
2 k_1 & -k_1 \\
-k_1 & 2 k_1
\end{bmatrix},
\]
respectively.
The covariances between the fluctuations of substrate and product concentrations, and the variances of the concentrations of each of these species, are found by solving Eq.~\eqref{Leq} at steady state:
\begin{align}
C_S^{\text{full}}&=\Omega^{-1} \frac{\phi_T \, x \big\{ 1 + \Gamma - x \Gamma + 
   \lambda \big[ 1 + x^2 + \Gamma - x (1 + \Gamma) \big] \big\}}
{(1+\lambda)\,\Gamma\,(1+\Gamma)}, \\
C_S^{\text{heur}}&=\Omega^{-1} \frac{\phi_T x}{\Gamma}, \\
C_P^{\text{full}}&=  
\frac{\Omega^{-1}\phi_T \, \lambda \, (x-1) x \big[ 
   \lambda (x-1) \Gamma (1+\Gamma) 
   - \big\{ (1+\Gamma)^2 + \lambda \big[ (x-1)x + (1+\Gamma)^2 \big] \big\} \Lambda 
   - (1+\lambda)(1+\Gamma)\Lambda^2
\big]}
{(1+\lambda)(1+\Gamma)\Lambda \big[\Lambda(1+\Gamma+\Lambda) 
   + \lambda \big(\Gamma - x \Gamma + \Lambda + \Gamma \Lambda + \Lambda^2 \big) \big]}, \\
C_P^{\text{heur}}&=\Omega^{-1}\frac{\phi_T \lambda (1-x)x}{(1+\lambda)\,\Lambda}, \\
C_{S,P}^{\text{full}}&=-\Omega^{-1}
\frac{\phi_T \, \lambda \, (x-1) x^2 \big[\lambda (x-1) - (1+\lambda)\Lambda \big]}
{(1+\lambda)(1+\Gamma)\big[ \lambda (x-1)\Gamma - (1+\lambda)(1+\Gamma)\Lambda - (1+\lambda)\Lambda^2 \big]}
,\quad\text{and} \\
C_{S,P}^{\text{heur}}&=0,
\end{align}
where $x = k_1/k_4\phi_T$, $\lambda = k_4/k_3$, $\Lambda = \tau_C/\tau_P$, and $\Gamma = \tau_C/\tau_S$. Note that $0 < x < 1$, see the main text, and that the timescales $\tau_i$ are defined in Eq.~\eqref{tscond}.

In the limit of timescale separation, i.e. for $\Lambda \rightarrow 0$ and $\Gamma \rightarrow 0$, and of infrequent catalysis ($\lambda \rightarrow 0$), taken at constant $x$ and $\phi_T$, we find that the variance and covariance predictions of the full and heuristic LNAs agree: $C_S^{\text{full}}/C_S^{\text{heur}} \rightarrow 1$, $C_P^{\text{full}}/C_P^{\text{heur}} \rightarrow 1$, and $C_{S,P} ^{\text{full}} - C_{S,P}^{\text{heur}} \rightarrow 0$ for all $x$.


The power spectrum matrices for the full and reduced systems are given by Eq.~\eqref{ps}:
\begin{align}
  \underline{P}^{\text{full}} (\omega) &= \Omega^{-1} (\underline{J}_{\text{full}} + i\underline{I}_3\omega)^{-1} \underline{D}_{\text{full}}\big(\underline{J}_{\text{full}}^T - i\underline{I}_3\omega\big)^{-1}, \\ 
  \underline{P}^{\text{heur}} (\omega) &= \Omega^{-1} (\underline{J}_{\text{heur}} + i\underline{I}_2\omega)^{-1} \underline{D}_{\text{heur}}\big(\underline{J}_{\text{heur}}^T - i\underline{I}_2\omega\big)^{-1};
\end{align}
by Eq.~\eqref{Coh}, the coherences between the substrate and product timeseries for the full and reduced systems read
\begin{align}
    \xi_{S,P}^{\text{full}}(\omega) &= \frac{|P_{12}^{\text{full}}(\omega)|^2}{P_{11}^{\text{full}}(\omega) P_{22}^{\text{full}}(\omega)} = \frac{c_0 + c_1 \omega^2 + c_2 \omega^4}{c_3 + c_4 \omega^2 + c_5 \omega^4 + c_6 \omega^6}\quad\text{and} \label {psfull} \\
    \xi_{S,P}^{\text{heur}}(\omega) &= \frac{|P_{12}^{\text{heur}}(\omega)|^2}{P_{11}^{\text{heur}}(\omega) P_{22}^{\text{heur}}(\omega)} = \frac{1}{4}, \label {psred}
\end{align}
where $c_i$ are constants that are defined as
\begin{align}
\begin{split}
c_0 &= k_3^6 \, \lambda^3 (1+\lambda)^3 \, \Gamma^2, \\
c_1 &= k_3^4 \, \lambda (1+\lambda) \big\{ 
        [1 + \lambda + 2\lambda (x-1)x]^2 
        + 2(1+\lambda)(1+\lambda x)\Gamma 
        + (1+\lambda)^2 \Gamma^2 \big\}, \\
c_2 &= k_3^2 \, \lambda (1+\lambda) \, \big(1 - 3x + 2x^2\big)^2, \\
c_3 &= 4 k_3^6 \, \lambda^3 (1+\lambda)^2 \, 
        [1 + \lambda + \lambda (x-1)x] \Gamma^2, \\
c_4 &= 4 k_3^4 \, \lambda (1+\lambda) \big(
        [1 + \lambda + \lambda (x-1)x]^2
        + 2(1+\lambda x)[1 + \lambda + \lambda (x-1)x]\Gamma + (1+\lambda)\{1 + \lambda[2 + x(2x-3)]\}\Gamma^2 
        \big), \\
c_5 &= 4 k_3^2 (x-1)^2 \big\{
        (1+2\lambda)[1 + \lambda + \lambda(x-1)x ] 
        + 2(1+\lambda)(1+\lambda x)\Gamma 
        + (1+\lambda)^2 \Gamma^2 \big\},\quad\text{and} \\
c_6 &= 4(x-1)^4.
\label{cs}
\end{split}
\end{align}
Since $0 < x < 1$, $\lambda > 0$, $k_3 > 0$, and $\Gamma > 0$, it follows that $c_i > 0$ for $i=0,1,3,4,5,6$, while $c_2$ is positive when $x \ne 1/2$ and equal to $0$ when $x = 1/2$. An important implication is that for the full LNA, in the limit of large $\omega$, $ \xi_{S,P}^{\text{full}}(\omega) \propto \omega^{-2}$ when $x \ne 1/2$ and $ \xi_{S,P}^{\text{full}}(\omega) \propto \omega^{-4}$ when $x = 1/2$. By contrast, $\xi_{S,P}^{\text{heur}}$ is independent of $\omega$. 

To understand the large discrepancy between the predictions of the full and heuristic LNAs, we consider the predicted power spectra of substrate and product fluctuations and their cross-spectrum, since these compose the coherence; see Eqs.~\eqref{psfull} and \eqref{psred}. We find that both the full LNA and the heuristic LNAs predict that the tails of the power spectra of substrate and product decay as $\omega^{-2}$, whereas the full LNA predicts that the tail of the squared modulus of the cross-spectrum of substrate and product decays as $\omega^{-6}$ while the heuristic LNA predicts a slower decay with $\omega^{-4}$, if $k_1 \ne k_4 \phi_T/2$. For the special case where $x = 1/2$, the squared modulus of the cross-spectrum in the full LNA decays as $\omega^{-8}$, while the heuristic LNA still predicts a slower decay at rate $\omega^{-4}$. Hence, the discrepancy between the coherences predicted by the two LNAs is due to discrepancies in the tails of the cross-spectra. 

\subsection{ssLNA}

We apply the ssLNA framework described in Section~\ref{rigorousred} to calculate the coherence spectrum for substrate and product, assuming that they are the slow species, while enzyme and complex are the fast species. 

The partitioned stoichiometric matrices of the slow and fast species are given by
\[
\underline{S}_s = 
\begin{bmatrix}
1 & -1 & 1 & 0 & 0 \\
0 & 0 & 0 & 1 & -1
\end{bmatrix}
\quad\text{and}\quad
\underline{S}_f = 
\begin{bmatrix}
0 & 1 & -1 & -1 & 0
\end{bmatrix}.
\] 
Note that $\underline{S}_s$ corresponds to the entries in the first and second rows of $\underline{S}_{\text{full}}$, while $\underline{S}_f$ equals the last row of $\underline{S}_{\text{full}}$ in Section~\ref{enzymeLNA}. 

The partitioned Jacobian matrices are given by
\begin{align*}
\underline{J}_{ss} =
\begin{bmatrix}
- k_2 \left( \phi_T - \dfrac{k_1}{k_4} \right) & 0 \\
0 & - k_5
\end{bmatrix}, \quad
& \underline{J}_{sf} =
\begin{bmatrix}
\dfrac{(k_1 + \phi_T k_3) k_4}{\phi_T k_4 - k_1} \\[1mm]
k_4
\end{bmatrix}, \\
\underline{J}_{fs} =
\begin{bmatrix}
k_2 \left( \phi_T - \dfrac{k_1}{k_4} \right) & 0
\end{bmatrix}, \quad\text{and}\quad
& \underline{J}_{ff} = - \dfrac{\phi_T\, k_4 (k_3 + k_4)}{\phi_T\, k_4 - k_1}.
\end{align*}
These matrices can be extracted from the Jacobian matrix $\underline{J}_{\text{full}}$ in Section~\ref{enzymeLNA}: $\underline{J}_{ss}$ is the upper left $2 \times 2$ sub-matrix; $\underline{J}_{sf}$ is the upper right $2 \times 1$ sub-matrix; $\underline{J}_{fs}$ is the lower left $1 \times 2$ sub-matrix; and $\underline{J}_{ff}$ is the lower right $1 \times 1$ sub-matrix, or scalar. Then, we can calculate the effective diffusion and Jacobian matrices as
\[
\underline{A} = \underline{S}_s \sqrt{\underline{F}} =
\begin{bmatrix}
\sqrt{k_1} & - \sqrt{\dfrac{k_1 (k_3 + k_4)}{k_4}} & \sqrt{\dfrac{k_1 k_3}{k_4}} & 0 & 0 \\
0 & 0 & 0 & \sqrt{k_1} & - \sqrt{k_1}
\end{bmatrix},
\]
\[
\underline{B} = \underline{J}_{sf} \underline{J}_{ff}^{-1} \underline{S}_f \sqrt{\underline{F}} =
\begin{bmatrix}
0 & - \dfrac{(k_1 + \phi_T k_3) \sqrt{\dfrac{k_1 (k_3 + k_4)}{k_4}}}{\phi_T (k_3 + k_4)} &
\dfrac{(k_1 + \phi_T k_3) \sqrt{\dfrac{k_1 k_3}{k_4}}}{\phi_T (k_3 + k_4)} &
\dfrac{\sqrt{k_1} (k_1 + \phi_T k_3)}{\phi_T (k_3 + k_4)} & 0 \\
0 & \dfrac{\sqrt{\dfrac{k_1 (k_3 + k_4)}{k_4}} (k_1 - \phi_T k_4)}{\phi_T (k_3 + k_4)} &
\dfrac{\sqrt{\dfrac{k_1 k_3}{k_4}} (\phi_T k_4 - k_1)}{\phi_T (k_3 + k_4)} &
\dfrac{\sqrt{k_1} (\phi_T k_4 - k_1)}{\phi_T (k_3 + k_4)} & 0
\end{bmatrix},
\]
\[
\underline{D}_S = (\underline{A}-\underline{B})(\underline{A}-\underline{B})^T =
\begin{bmatrix}
\dfrac{2 k_1 (k_1^2 - \phi_T k_1 k_4 + \phi_T^2 k_4 (k_3 + k_4))}{\phi_T^2 k_4 (k_3 + k_4)} &
- \dfrac{k_1 (2 k_1^2 - 2 \phi_T k_1 k_4 + \phi_T^2 k_4 (k_3 + k_4))}{\phi_T^2 k_4 (k_3 + k_4)} \\
- \dfrac{k_1 (2 k_1^2 - 2 \phi_T k_1 k_4 + \phi_T^2 k_4 (k_3 + k_4))}{\phi_T^2 k_4 (k_3 + k_4)} &
\dfrac{2 k_1 (k_1^2 - \phi_T k_1 k_4 + \phi_T^2 k_4 (k_3 + k_4))}{\phi_T^2 k_4 (k_3 + k_4)}
\end{bmatrix},
\]
and
\[
\underline{J}_S = \underline{J}_{ss} - \underline{J}_{sf} \underline{J}_{ff}^{-1} \underline{J}_{fs} =
\begin{bmatrix}
- \dfrac{(k_1 - \phi_T k_4)^2}{\phi_T k_4 K_M} & 0 \\
\dfrac{(k_1 - \phi_T k_4)^2}{\phi_T k_4 K_M} & - k_5
\end{bmatrix}.
\]
Note that $\underline{F}$ is a diagonal matrix with entries corresponding to the components of the vector $\vec{f}_{\text{full}}$, recall Section~\ref{enzymeLNA}.
The covariance between the fluctuations of substrate and product concentrations, and the variances of the concentrations of each of these species, are found from Eq.~\eqref{Leq} with $\underline{J} = \underline{J}_S$ and $\underline{D} = \underline{D}_S$ at steady state:
\begin{align}
C_{S}^{\text{ssLNA}} &= \frac{\phi_T \, x \, [ 1 + \lambda + \lambda (x-1) x ]}{(1 + \lambda) \, \Gamma}, \\[2mm]
C_{S,P}^{\text{ssLNA}} &= \frac{\phi_T \, \lambda^2 \, (x-1)^2 x^2}{(1 + \lambda) \, [\Lambda + \lambda (\Gamma - x \Gamma + \Lambda)]},\quad\text{and} \\[1mm]
C_{P,P}^{\text{ssLNA}} &= \frac{\phi_T \, \lambda \, (1 - x) x \, \{ \Lambda + \lambda [\Gamma - x \Gamma + \Lambda + (x-1)x \Lambda] \}}{(1 + \lambda) \, \Lambda \, [\Lambda + \lambda (\Gamma - x \Gamma + \Lambda) ]}.
\end{align}
In the limit of timescale separation, i.e. for $\Lambda \rightarrow 0$ and $\Gamma \rightarrow 0$, taken at constant $x$, $\phi_T$, and $\lambda$, we find that the variances and covariances predicted by the full LNA and the ssLNA agree: $C_S^{\text{full}}/C_S^{\text{ssLNA}} \rightarrow 1$, $C_P^{\text{full}}/C_P^{\text{ssLNA}} \rightarrow 1$, and $C_{S,P}^{\text{full}}/C_{S,P}^{\text{ssLNA}} \rightarrow 1$ for all $x$.

The power spectrum matrix is found from Eq.~\eqref{ps}:
\begin{align}
  \underline{P}^{\text{ssLNA}} (\omega) &= \Omega^{-1} \big(\underline{J}_S + i\underline{I}_2\omega\big)^{-1} \underline{D}_S\big(\underline{J}_S^T - i\underline{I}_2\omega\big)^{-1}.
\end{align}
The coherence between the substrate and product timeseries is then given by Eq.~\eqref{Coh},
\begin{align}
    \xi_{S,P}^{\text{ssLNA}}(\omega) &= \frac{|P_{12}^{\text{ssLNA}}(\omega)|^2}{P_{11}^{\text{ssLNA}}(\omega) P_{22}^{\text{ssLNA}}(\omega)} = \frac{d_0 + d_1 \omega^2}{d_2 + d_3 \omega^2},
\end{align}
where
\begin{align}
\begin{split}
d_0 &= \phi_T^6 \, k_3^6 \, \lambda^4 \, (1 + \lambda)^2 \, (x-1)^4, \\
d_1 &= \phi_T^4 \, k_3^4 \, K_M^2 \, \lambda^2 \, \big[ 1 + \lambda + 2 \lambda (x-1) x \big]^2, \\
d_2 &= 4 \, \phi_T^6 \, k_3^6 \, \lambda^4 \, (1 + \lambda) \, (x-1)^4 \, \big[1 +\lambda (1 + (x-1)x) \big],\quad\text{and} \\
d_3 &= 4 \, \phi_T^4 \, k_3^4 \, K_M^2 \, \lambda^2 \, \big[ 1 + \lambda + \lambda (x-1) x \big]^2.
\label{ds}
\end{split}
\end{align}
Note that $d_i > 0$ for $i=0,\dots,3$, as well as that $d_0/d_2 = c_0/c_3$ in Eq.~\eqref{cs}; hence, the coherence predictions of the ssLNA and the full LNA agree exactly at $\omega = 0$. In the limit of large $\omega$, the coherence of the ssLNA tends to the constant $d_1/d_3$. Applying the limit of inefficient catalysis, with $\lambda \rightarrow 0$, we obtain $d_1/d_3 \rightarrow 1/4$, which is the value predicted by the heuristic LNA in Eq.~\eqref{psred}. Hence, the ssLNA and the heuristic LNA agree for large frequencies provided that there is timescale separation, as assumed by the ssLNA, and that the additional assumption of inefficient catalysis holds. 

\section{Calculations for simple model of translation} \label{AppC}

The stoichiometric matrix and rate function vector for the reaction scheme in Eq.~\eqref{transmodelfull} are given by
\[
\underline{S}_{\text{full}} =
\begin{bmatrix}
  1 & -1 & -1 & 0 & 1 & 0 \\
  0 & 0  & 0  & 1 & 0 & -1 \\
  0 & 0  & 1  & 0 & -1 & 0
\end{bmatrix}
\quad\text{and}\quad
\vec{f}_{\text{full}} =
\begin{bmatrix}
  k_{1} \\
  k_{2}\,\phi_{M} \\
  k_{3}\,\phi_{M} \\
  k_{4}\,C \\
  k_{5}\,C \\
  k_{6}\,\phi_{P}
\end{bmatrix}.
\]
Note that the element in the first row and $j$-th column of the matrix $\underline{S}_{\text{full}}$ represents the net change in the number of molecules of $M$ when the $j$-th reaction, associated with reaction rate constant $k_j$, occurs. Similarly, the second and third rows describe the changes in the numbers of molecules of $P$ and $C$, respectively. 

For the heuristically reduced reaction scheme, Eq.~\eqref{transmodelred}, we can write
\[
\underline{S}_{\text{heur}} =
\begin{bmatrix}
  1 & -1 & 0 & 0 \\
  0 & 0  & 1 & -1
\end{bmatrix}
\quad\text{and}\quad
\vec{f}_{\text{heur}} =
\begin{bmatrix}
  k_{1} \\
  k_{2}\,\phi_{M} \\
  k'\phi_{M} \\
  k_{6}\,\phi_{P}
\end{bmatrix}.
\]
The first and second rows in the stoichiometric matrix $\underline{S}_{\text{heur}}$ describe the changes in the numbers of molecules of $M$ and $P$, respectively. The $i$-th column is associated with the reaction whose rate is given by the $i$-th entry of $\vec{f}_{\text{heur}}$.

The rate equations, Jacobian and diffusion matrices, Lyapunov equations, power spectra, and coherence spectra for the full LNA can be constructed by replacing $\underline{S}$ and $\vec{f}$ by $\underline{S}_{\text{full}}$ and $\vec{f}_{\text{full}}$, respectively, in the equations in Section~\ref{LNA_MI}. Similarly, the heuristic LNA can be constructed by replacing $\underline{S}$ by $\underline{S}_{\text{heur}}$ and $\vec{f}$ by $\vec{f}_{\text{heur}}$ in Section \ref{LNA_MI}.

For the ssLNA, the relevant matrices read
\[
\underline{S}_{s} =
\begin{bmatrix}
  1 & -1 & -1 & 0 & 1 & 0 \\
  0 & 0  & 0  & 1 & 0 & -1
\end{bmatrix}
\quad\text{and}\quad
\underline{S}_{f} =
\begin{bmatrix}
  0 & 0 & 1 & 0 & -1 & 0
\end{bmatrix},
\]
as well as
\[
\underline{J}_{ss} =
\begin{bmatrix}
  -k_{2} - k_{3} & 0 \\
  0 & -k_{6}
\end{bmatrix},
\quad
\underline{J}_{sf} =
\begin{bmatrix}
  k_{5} \\
  k_{4}
\end{bmatrix},\quad
\underline{J}_{fs} =
\begin{bmatrix}
  k_{3} & 0
\end{bmatrix},
\quad\text{and}\quad
\underline{J}_{ff} = -k_{5}.
\]
Note that $\underline{S}_s$ and $\underline{S}_f$ correspond to the first two rows and to the last row, respectively, of the matrix $\underline{S}_{\text{full}}$. The matrices $\underline{J}_{ss}$, $\underline{J}_{sf}$, $\underline{J}_{fs}$, and $\underline{J}_{ff}$ are the block sub-matrices of the Jacobian matrix of the full system, Eq.~\eqref{Jac}, with $\underline{S}$ and $\vec{f}$ replaced by $\underline{S}_{\text{full}}$ and $\vec{f}_{\text{full}}$, respectively. Substituting the matrices $\underline{S}_{s}$, $\underline{S}_{f}$, $\underline{J}_{ss}$, $\underline{J}_{sf}$, $\underline{J}_{fs}$, and $\underline{J}_{ff}$ into the relevant equations of Section~\ref{rigorousred}, one obtains the effective Jacobian and diffusion matrices from which one can construct and solve the equations for the covariance matrix, power spectra, and coherence spectra of the ssLNA. 

Solving the Lyapunov equations of the full LNA, the heuristic LNA, and the ssLNA at steady state, one obtains the following relationships between the corresponding second moments:

\begin{align}
\frac{C_{M}^{\text{full}}}{C_{M}^{\text{heur}}} &= 1, 
~\frac{C_{M,P}^{\text{full}}}{C_{M,P}^{\text{heur}}} = 
\frac{\beta + \gamma}{\beta (\alpha + \beta + \gamma + 1) + \gamma}, \\[6pt]
\frac{C_{P}^{\text{full}}}{C_{P}^{\text{heur}}} &= 
\frac{(\beta + \gamma)\, [\lambda (\alpha + \beta + \gamma) + \beta (\alpha + \beta + \gamma + 1) + \gamma]}{[\beta (\alpha + \beta + \gamma + 1) + \gamma]\, (\alpha \lambda + \beta + \gamma)}, \\[6pt]
\frac{C_{M}^{\text{full}}}{C_{M}^{\text{ssLNA}}} &= 1, 
~\frac{C_{M,P}^{\text{full}}}{C_{M,P}^{\text{ssLNA}}} = 
\frac{\beta + \gamma}{\beta (\alpha + \beta + \gamma + 1) + \gamma},\quad\text{and} \\[6pt]
\frac{C_{P}^{\text{full}}}{C_{P}^{\text{ssLNA}}} &= 
\frac{(\beta + \gamma)\, [\lambda (\alpha + \beta + \gamma) + \beta (\alpha + \beta + \gamma + 1) + \gamma]}{[\beta (\alpha + \beta + \gamma + 1) + \gamma]\, [\lambda (\alpha + \beta + \gamma) + \beta + \gamma]},
\end{align}
where $\alpha = k_3/k_5$, $\beta = k_6/k_5$, $\lambda = k_4/k_5$, and $\gamma = k_2/k_5$. Note that $\alpha+\gamma = \tau_c/\tau_M$ and $\beta = \tau_C/\tau_P$ represent the ratios of the timescale of complex to the timescales of mRNA and protein, respectively. 

\section{Calculations for model of translation with compartmentalization and transcriptional bursting} \label{AppD}

The stoichiometric matrix and rate function vector for the reaction scheme in Eq.~\eqref{transmodelfull1} are given by
\[
\underline{S}_{\text{full}} =
\begin{bmatrix}
0 & 0 & 1 & -1 & 0 & 0 \\
0 & 0 & 0 & 0 & 1 & -1 \\
r & -1 & -1 & 0 & 0 & 0
\end{bmatrix}
\quad\text{and}\quad
\vec{f}_{\text{full}} =
\begin{bmatrix}
k_1 \\
k_2 M_{n} \\
k_3 M_{n} \\
k_4 M_c \\
k_5 M_c \\
k_6 P
\end{bmatrix},
\]
respectively. The element in the first row and $j$-th column of $\underline{S}_{\text{full}}$ represents the net change in the number of molecules of $M$ when the $j$-th reaction, associated with reaction rate constant $k_j$, occurs. Similarly, the second and third rows describe the changes in the numbers of molecules of $P$ and $M_n$, respectively. The integer $r$ is sampled randomly from the geometric distribution with support on the non-negative integers and mean $b$.

For the reduced reaction scheme, Eq.~\eqref{transmodelred1}, we have
\[
\underline{S}_{\text{heur}} =
\begin{bmatrix}
r & -1 & 0 & 0 \\
0 & 0 & 1 & -1
\end{bmatrix}
\quad\text{and}\quad
\vec{f}_{\text{heur}} =
\begin{bmatrix}
k' \\
k_4 M_c \\
k_5 M_c \\
k_6 P
\end{bmatrix},
\]
where $k' = k_1 k_3/(k_2 + k_3)$. The first and second rows in $\underline{S}_{\text{heur}}$ describe the changes in the numbers of molecules of $M_c$ and $P$, respectively. The $i$-th column is associated with the reaction whose rate is given by the $i$-th entry of $\vec{f}_{\text{heur}}$.

The rate equations, Jacobian and diffusion matrices, Lyapunov equations, power spectra, and coherence spectra for the full LNA can be constructed by replacing $\underline{S}$ by $\underline{S}_{\text{full}}$ and $\vec{f}$ by $\vec{f}_{\text{full}}$ in the equations in Section~\ref{LNA_MI}. Similarly, the heuristic LNA can be constructed by replacing $\underline{S}$ and $\vec{f}$ by $\underline{S}_{\text{heur}}$ and $\vec{f}_{\text{heur}}$, respectively, in Section~\ref{LNA_MI}. Note that wherever a factor of $r^2$ appears in the diffusion matrices, one should replace it by $\langle r^2 \rangle$, which is the mean square burst size calculated from the geometric distribution with mean burst size $b$: $\langle r^2 \rangle = \sum_{k=0}^\infty k^2(1 - p)^k p = b (1 + 2b)$, where $p = 1/(1+b)$. Note also that $\langle r \rangle = b$. This averaging over powers of $r$ in the diffusion matrices is necessary, given that $r$ is not fixed but, rather, a randomly sampled integer. 

For the ssLNA, the relevant matrices are given by
\[
\underline{S}_{s} =
\begin{bmatrix}
0 & 0 & 1 & -1 & 0 & 0 \\
0 & 0 & 0 & 0 & 1 & -1
\end{bmatrix}
\quad\text{and}\quad
\underline{S}_{f} =
\begin{bmatrix}
r & -1 & -1 & 0 & 0 & 0
\end{bmatrix},
\]
as well as by
\[
\underline{J}_{ss} =
\begin{bmatrix}
- k_4 & 0 \\
k_5 & - k_6
\end{bmatrix},
\quad
\underline{J}_{sf} =
\begin{bmatrix}
k_3 \\
0
\end{bmatrix}, \quad
\underline{J}_{fs} =
\begin{bmatrix}
0 & 0
\end{bmatrix},
\quad\text{and}\quad
\underline{J}_{ff} = -k_{2}-k_{3}.
\]
Note that $\underline{S}_s$ and $\underline{S}_f$ correspond to the first two rows and to the last row, respectively, of the matrix $\underline{S}_{\text{full}}$. The matrices $\underline{J}_{ss}$, $\underline{J}_{sf}$, $\underline{J}_{fs}$, and $\underline{J}_{ff}$ are the block sub-matrices of the Jacobian matrix of the full system, Eq.~\eqref{Jac}, with $\underline{S}$ and $\vec{f}$ replaced by $\underline{S}_{\text{full}}$ and $\vec{f}_{\text{full}}$, respectively. Substituting the matrices $\underline{S}_{s}$, $\underline{S}_{f}$, $\underline{J}_{ss}$, $\underline{J}_{sf}$, $\underline{J}_{fs}$, and $\underline{J}_{ff}$ into the relevant equations of Section~\ref{rigorousred}, one obtains the effective Jacobian and diffusion matrices from which one can construct and solve the equations for the covariance matrix, power spectra, and coherence spectra of the ssLNA. As before, wherever a factor of $r^2$ appears in the diffusion matrices, one should replace it by $\langle r^2 \rangle$.

\bibliographystyle{apsrev}
\bibliography{ref}

\begin{thebibliography}{52}
\bibitem[{myf()}]{myfootnote}
\bibinfo{note}{The Gillespie algorithm samples paths from the CME of the reduced system, Eq.~\eqref{redenzyme}, which has propensities $k_1 \Omega$ for the reaction $\varnothing \xrightarrow{k_1} S$, $k' n_S = \frac{k_4 \phi_T n_S}{\Omega K_M + n_S}$ for the reaction $S \xrightarrow{k'} P$, and $k_5 n_p$ for the reaction $P\xrightarrow{k_5}\varnothing$; here, $\Omega$ is the volume of the system and $n_i$ denotes the number of molecules of species $i$.}
\end{thebibliography}

\end{document}